\documentclass[%
superscriptaddress,
showpacs,preprintnumbers,
nofootinbib,
amsmath,amssymb,
aps,
prd,
tightenlines,
11pt,
notitlepage
]{revtex4-1}
\usepackage[T1]{fontenc}

\usepackage{alltt}
\usepackage{amsmath}
\usepackage{amsthm}
\usepackage{amssymb}
\usepackage{bm}
\usepackage{booktabs,siunitx}
\usepackage{cancel}
\usepackage{color}
\usepackage{csquotes}
\usepackage{enumerate}
\usepackage{graphicx,bm}
\usepackage{lipsum}
\usepackage{listings}
\usepackage{mathtools}
\usepackage{mathrsfs}
\usepackage{physics}
\usepackage{relsize}
\usepackage{slashed}
\usepackage{soul}
\usepackage{spverbatim}
\usepackage{wrapfig}
\usepackage{xcolor}
\usepackage{diagbox}
\usepackage{multirow}
\usepackage{subcaption}

\usepackage[utf8]{inputenc}

\usepackage[pdftex,hypertexnames=false,linktocpage=true]{hyperref}
\hypersetup{colorlinks=true,linkcolor=blue,anchorcolor=blue,citecolor=darkgreen,filecolor=blue,urlcolor=blue,bookmarksnumbered=true,
	pdfview=FitB
}
\colorlet{darkgreen}{green!60!black}
\colorlet{brightyellow}{yellow!75!red}
\colorlet{orange}{red!50!yellow}
\colorlet{darkblue}{blue!60!black}
\colorlet{darkred}{red!80!black}
\colorlet{greenblue}{green!50!blue}

\makeatletter

\newcommand{\Rmnum}[1]{\expandafter\@slowromancap\romannumeral #1@}
\makeatother

\def\dd{{\mathrm{d}}}
\def\imag{{\mathrm{i}}}

\begin{document}
	
	\title{Heavy-Light Mesons on the Light Front}
	
	\author{Shuo Tang}
	%\email{tang@iastate.edu}	
	\affiliation{%
		Department of Physics and Astronomy, Iowa State University, Ames, IA 50011, USA
	}%
	
	\author{Yang Li}
	\email[Corresponding author: ] {leeyoung@iastate.edu}
%	\affiliation{%
%		Hebei Key Laboratory of Compact Fusion, Langfang 065001, China}
%	\affiliation{%
%		ENN Science and Technology Development Co., Ltd., Langfang 065001, China
%	}	
	\affiliation{%
		Department of Physics and Astronomy, Iowa State University, Ames, IA 50011, USA
	}	\affiliation{%
	School of Nuclear Science and Technology, University of Chinese Academy of Sciences, Beijing 100049, China
}
	
	\author{Pieter Maris}
	\affiliation{%
		Department of Physics and Astronomy, Iowa State University, Ames, IA 50011, USA
	}
	
	\author{James P. Vary}
	\affiliation{%
		Department of Physics and Astronomy, Iowa State University, Ames, IA 50011, USA
	}
	%\collaboration{MUSO Collaboration}%\noaffiliation
	
	%\author{Charlie Author}
	%\homepage{http://www.Second.institution.edu/~Charlie.Author}
	%\affiliation{
	%	Second institution and/or address\\
	%	This line break forced% with \\
	%}%
	%\affiliation{
	%	Third institution, the second for Charlie Author
	%}%
	%\author{Delta Author}
	%\affiliation{%
	%	Authors' institution and/or address\\
	%	This line break forced with \textbackslash\textbackslash
	%}%
	
	%\collaboration{CLEO Collaboration}%\noaffiliation
	
	\date{\today}% It is always \today,
	%  but any date may be explicitly specified
	
	\begin{abstract}
		We study the heavy-light mesons within basis light-front quantization. The resulting mass spectra of $D$, $D_s$, $B$, and $B_s$ agree reasonably well with experiments. We also predict states which could be measured in the near future. In the light-front formalism, we calculate the light-front wave functions and additional experimental observables, such as parton distribution functions, distribution amplitudes, and decay constants by means of integrations over light-front wave functions. 
		We also provide ratios of decay constants for selected pseudoscalar meson decays ($D_s$ to $D$ and $B_s$ to $B$) as they may prove to be theoretically more robust and more reliably determined in experiments.
		We find that our ratios are systematically smaller than existing experiment and other approaches by $5-18\%$.
	\end{abstract}
	\maketitle
	
	\section{Introduction}

The motivation for studying heavy-light mesons  ($D$, $D_s$, $B$, $B_s$) with basis light-front quantization (BLFQ)~\cite{PhysRevC.81.035205} is twofold. First, the combination of a heavy and a light quark is the closest QCD analogue of the hydrogen atom in QED, so that similarities and differences in spectroscopic features could inform discussions of the relative roles of gauge-boson exchange and confinement. Second, the successful applications of BLFQ to heavy meson systems~\cite{PhysRevD.96.016022,PhysRevD.98.114038} provides a foundation for understanding the roles of key elements of the quark-antiquark effective Hamiltonian adopted for the heavy-light system. We employ the same form of Hamiltonian with only two fit parameters in order to test the validity of our model as well as some novel behavior for the heavy-light mesons.

%		While the spectroscopy of the heavy-light mesons has been inspected, a number of interesting issues remain. For instance, 
Heavy-light systems have been studied widely by various methods for a long time, and different approaches emphasize certain features of the system and have their own advantages. Examples include the heavy-quark effective theory~\cite{ISGUR1989113};
%, an effective theory characterizing the heavy quark symmetry; 
models that incorporate chiral dynamics owing to the light constituent quark~\cite{PhysRevD.48.4370,PhysRevD.49.409}; QCD sum rules~\cite{NIELSEN201041}; Dyson-Schwinger equation approach~\cite{MARIS2006136}; etc. Nevertheless, there are detailed properties of the excitation spectrum of heavy-light mesons that are yet to be fully understood. In addition, recent discoveries of new excited states, including the first heavy flavored spin-3 resonance $D^*_{s3}(2860)^\pm$ state, and $B_J(5840)^{0,+},\ B_J(5960)^{0,+}$~\cite{PhysRevLett.113.162001,Aaij2015} have stimulated renewed quests for theoretical understandings. 
%It provides great opportunities for the phenomenology to test and improve their theory, and then predict more hadron states with higher accuracy.

Additional motivation stems from the fact that the leptonic decay constant of heavy-light mesons plays a significant role in determining the Cabibbo-Kobayashi-Maskawa (CKM) quark-mixing matrix elements within Standard Model~\cite{PhysRevLett.10.531,10.1143/PTP.49.652}. Moreover, with continually improving experimental precision, these observables constitute powerful tools to probe physics beyond Standard Model~\cite{NARISON2008308}. 

%Heavy-light systems have been studied widely by various methods for a long time, and different approaches emphasize certain features of the system and have their own advantages. Examples include the heavy quark effective theory \cite{ISGUR1989113}, an effective theory characterizing the heavy quark symmetry; Models incorporate chiral dynamics owing to the light constituent quark \cite{PhysRevD.48.4370,PhysRevD.49.409}; QCD sum rules \cite{NIELSEN201041}, and etc.
%
In this paper, we discuss the heavy-light mesons within BLFQ, a non-perturbative Hamiltonian framework. The paper is organized as follows. Sec.~\ref{sec1} outlines the elements of BLFQ, the theoretical approach to the heavy-light systems. Then in Sec.~\ref{sec2}, we exhibit the numerical results of the mass spectrum, light-front wave functions, parton distribution functions, distribution amplitudes, and the decay constants.  Sec.~\ref{sec3} provides our discussions and outlook.

\section{Basis Light-Front quantization for Heavy-Light Systems}
\label{sec1}

In this work, we adopt the effective Hamiltonian within the $\ket{q\bar{q}}$ Fock sector in the form introduced for the heavy mesons~\cite{PhysRevD.96.016022,PhysRevD.98.114038}. It comprises two parts $H_\text{tot} = H_0 + V^\text{eff}_g$, where
\begin{equation}
\begin{aligned}
H_0 = \frac{\vec{k}^2_\perp+m_q^2}{x} + \frac{\vec{k}^2_\perp+m_{\bar{q}}^2}{1-x} + \kappa^4\vec{\zeta}^2_\perp - \frac{\kappa^4}{(m_q+m_{\bar{q}})^2} \partial_x (x(1-x)\partial_x),
\end{aligned}
\label{eq3}
\end{equation}
is the holographic QCD Hamiltonian~\cite{PhysRevLett.102.081601,LI2016118} augmented by massive quark kinematics and the longitudinal confinement~\cite{PhysRevD.96.016022,PhysRevD.98.114038}. In addition, the spin structure of the hadrons is generated by the effective one-gluon-exchange potential $V^\text{eff}_g$. Along with a running coupling $\alpha_s(Q^2)$, $V^\text{eff}_g$ governs the short-range physics:
\begin{equation}
V^\text{eff}_g = -\frac{C_F 4 \pi \alpha_s(Q^2)}{Q^2} \bar{u}_{s'}(k')\gamma_\mu  u_s(k) \bar{v}_{\bar{s}}(\bar{k}) \gamma^\mu v_{\bar{s}'}(\bar{k}').
\end{equation}
With the Hamiltonian, one can solve the light-front eigenvalue problem $H_\text{tot} \ket{\Psi_h(P,J,m_J)} = M^2 \ket{\Psi_h(P,J,m_J)} $ for the bound state eigenmasses and light-front wave functions (LFWFs). In the BLFQ framework, we expand the LFWFs in terms of basis functions for the transverse and longitudinal directions, respectively:
\begin{equation}
\label{eq1}
\begin{aligned}
\psi_{s\bar{s}/h}^{(m_J)} (x, \vec{k}_\perp) = \sum_{n,m,l} \psi_h(n,m,l,s,\bar{s}) 
\phi_{nm} \big(k_\perp/\sqrt{x(1-x)}\big) \chi_l(x),
\end{aligned}
\end{equation}
where $k_\perp \equiv \abs{\vec{k}_\perp}$. The transverse basis functions,
\begin{equation}
\begin{aligned}
\phi_{nm} (\vec{q}_\perp) = \frac{1}{b}\sqrt{\frac{4\pi n!}{(n+\abs{m})!}} \Big(\frac{q_\perp}{b}\Big) ^{\abs{m}} e^{-\frac{1}{2} q^2_\perp / b^2} 
 L_n^{\abs{m}} (q^2_\perp/b^2) e^{\imag m \theta_q},
\end{aligned}
\label{eq4}
\end{equation}
are the 2D harmonic oscillator functions where $n$ and $m$ are the principal and orbital quantum numbers, respectively; $\vec{q}_\perp = \vec{k}_\perp/\sqrt{x(1-x)}$, with $x$ being the longitudinal momentum fraction carried by the quark $x=p^+_q/(p^+_q+ p^+_{\bar{q}})$; $\theta_q = \arg \vec{q}_\perp$, $b$ sets the scale of the basis, and  $L_n^{\abs{m}}$ is the associated Laguerre polynomial. The longitudinal basis functions are given by
\begin{equation}
\begin{aligned}
\chi_l(x) = \sqrt{4\pi(2l+\alpha+\beta+1)}\sqrt{\frac{\Gamma(l+1)\Gamma(l+\alpha+\beta+1)}{\Gamma(l+\alpha+1)\Gamma(l+\beta+1)}}  x^{\frac{\beta}{2}}(1-x)^{\frac{\alpha}{2}} P_l^{(\alpha,\beta)}(2x-1),\\
\end{aligned}
\end{equation}
where $P_l^{(\alpha,\beta)}(2x-1)$ is the Jacobi polynomial with quantum number $l$; $\alpha$ and $\beta$ are two dimensionless parameters associated with the constituent (anti-) quark masses~\cite{LI2016118}:
\begin{equation}
\alpha = 2m_{\bar{q}} (m_q + m_{\bar{q}} )/\kappa; \qquad \beta = 2m_q (m_q + m_{\bar{q}} )/\kappa.
\end{equation}
In practical calculations, we truncate the infinite basis by restricting the quantum numbers:
\begin{equation}
2n+\abs{m} + 1 \le N_\text{max}, \qquad 0 \le l \le L_\text{max}.
\end{equation}
Due to the transverse cutoff $N_\text{max}$, the ultraviolet (UV) and infrared (IR) regulators are implicit through $\Lambda_\text{UV} \approx b\sqrt{N_\text{max}}$, and $\Lambda_\text{IR} \approx b/\sqrt{N_\text{max}}$, respectively.  $L_\text{max}$ represents the basis resolution in the longitudinal direction. In this work, we choose $N_\text{max} = L_\text{max} $ for simplicity.
Notice that the total angular momentum projection is defined by $m_J = m + s +\bar{s}$, which is conserved in the constructed basis.
Solving the eigenequation by diagonalizing the $H_\text{tot}$ matrix, one obtains the eigenvalues which indicate the spectra as squared masses, and the eigenvectors which are the coefficients $ \psi_h(n,m,l,s,\bar{s}) $ in Eq.~\eqref{eq1}.

\section{Numerical Results}
\label{sec2}

In the framework of BLFQ, the confining strength and constituent (anti-) quark masses are typically taken as free parameters that are fitted to experiment.
Following the practice adopted for the heavy quarkonia systems~\cite{PhysRevD.98.114038,LI2016118}, we take the confining strength of longitudinal and transverse directions to be the same as reflected in Eq.~\eqref{eq3}, and we set the 2D harmonic oscillator scale parameter by $ b = \kappa$.
We adopt the running coupling as regularized in Ref.~\cite{LI2016118} now with the flavor number ($N_f$) dependence listed in Table. \ref{tb1}. Our convention is to set the heavier quarks, charm ($\mathsf{c}$) and bottom ($\mathsf{b}$) to be the quarks, while setting the lighter quarks, up ($\mathsf{u}$), down ($\mathsf{d}$), and  strange ($\mathsf{s}$) to be the antiquarks in the heavy-light systems. We take the masses of $\mathsf{u}$ and $\mathsf{d}$ antiquarks to be the same. Unless otherwise stated, we use $B$ to denote $B^0$ and $B^-$, and $D$ for $D^0$ and $D^+$.

Following the scheme adopted in Ref.~\cite{PhysRevD.98.114038}, the confining strength of the flavored meson is implemented as $\kappa_{\mathsf{x}\bar{\mathsf{y}}} = \sqrt{(\kappa^2_{\mathsf{x}\bar{\mathsf{x}}}+\kappa^2_{\mathsf{y}\bar{\mathsf{y}}})/2}$. We adopt $\kappa_{\mathsf{u}\bar{\mathsf{u}}/\mathsf{d}\bar{\mathsf{d}}} = \kappa_{\mathsf{s}\bar{\mathsf{s}}} = 0.59\text{ GeV}$~\footnote{We have tested our model with another confining strength $\kappa = 0.54$ GeV provided by Ref.~\cite{BRODSKY20151}, and found the overall spectra were not significantly affected: the deviation is no more than $20\%$ in r.m.s or $6\%$ in $\overline{\delta_J M} $ through the four heavy-light systems presented in Table~\ref{tb1}.}
as the confining strength for the light mesons~\cite{BRODSKY20151}, while the values for heavy quarkonia, $\kappa_{\mathsf{b}\bar{\mathsf{b}}} $ and $\kappa_{\mathsf{c}\bar{\mathsf{c}}} $, are taken from Ref.~\cite{PhysRevD.96.016022}.  Following our previous works on heavy systems~\cite{PhysRevD.96.016022,PhysRevD.98.114038}, we keep the same heavy quark masses ($m_\mathsf{b}$, $m_\mathsf{c}$), while we fit the light-quark masses ($m_\mathsf{u}/m_\mathsf{d}$, $m_\mathsf{s}$) to reproduce the heavy-light meson mass spectra from experiments. 
%More specifically, we tune the $m_\mathsf{u}\ (m_\mathsf{s})$ by comparing the mass of the first two low-lying states of $D$ and $B$ ($D_s$ and $B_s$), with $D^0, \ D^*(2007)$ and $B^\pm, \ B^*$ ($D_s^\pm,\  D_s^{*\pm}$ and $B_s^0,\  B_s^{*0}$) that measured experimentally, to achieve the least r.m.s. deviation. 
More specifically, we tune only the $m_\mathsf{u}/m_\mathsf{d}\ (m_\mathsf{s})$ to minimize the r.m.s. mass deviation between experiment and theory for the lowest pseudoscalar and vector states, $D^0$, $D^*(2007)$, $B^\pm$, $B^*$ ($D_s^\pm,\  D_s^{*\pm}$, $B_s^0,\  B_s^{*0}$).
The $m_\mathsf{u}/m_\mathsf{d}$ quark mass we obtain agrees with the value fitted directly to the light mesons excluding the pion~\cite{WenyangQian}.

The model parameters of heavy-light mesons are listed in Table~\ref{tb1}. We will elaborate on the details of our heavy-light mass spectra in the following section.
While heavy (e.g. $\eta_c$) and light mesons (e.g. $\rho, \pi$) are not the subject of interests in this work, we nevertheless include the obtained values for comparison purpose only. 
%where we also include the heavy and light mesons for comparison. 
%	It is worth noticing that the spectrum of light mesons are calculated by using $m_\mathsf{u}$ and $m_\mathsf{s}$ fitted from heavy-light mesons.
%	When we apply our fitted masses to the light mesons with the confining strength taken to be 0.59 GeV, we obtain the pion and rho meson masses at 338 MeV and 1121 MeV; the $K $ and  $K^*$ at 548 MeV  and 1307 MeV. Due 
Note that due to our omission of chiral dynamics, these obtained light meson masses differ substantially from the experimental measurement, as expected.
Within the same framework, Ref.~\cite{PhysRevC.99.035206} shows that incorporation a NJL type chiral symmetry breaking interaction dramatically improves the predictions for $\rho$, $\pi$ masses and additional observables.

%	One can notice that, for light mesonrate results on mass because this model did not include the chiral effect. However, it does not influence the heavier systems, even the one contains a light antiquark. We employ the 

\begin{table*}
	\centering 
	\begin{tabular}{ccccccccccc}
		\hline\noalign{\smallskip}
		&	\multirow{2}{*}{  $N_f$ }	& \multirow{2}{*}{ $m_q$ (GeV) } 	& \multirow{2}{*}{ $m_{\bar{q}}$ (GeV) } & \multirow{2}{*}{ $\kappa$ (GeV) }  & \multicolumn{2}{c}{\underline{ \hspace{0.1cm}Mass (GeV) \hspace{0.1cm} }}&    \hspace{0.1cm} r.m.s.(MeV)   \hspace{0.1cm}  &  \multirow{2}{*}{ $\overline{\delta_J M}$ (MeV) }\\
		&	&	&	&	&  \hspace{0.1cm} PS & V 	& $[N_\text{exp}]$	& \\
		\noalign{\smallskip}\hline\noalign{\smallskip}
		$\mathsf{u}\bar{\mathsf{u}}/\mathsf{d}\bar{\mathsf{d}}$		& 3		&0.553			&0.553		& 0.590			& 0.338 		& 1.121			& ...					& 36 (3.3\%)	\\
		$\mathsf{s}\bar{\mathsf{u}}$												& 3		&0.647			&0.553		& 0.590			&0.548	 		& 1.307 		& ... 					& 30 (2.5\%)	\\
		$\mathsf{s}\bar{\mathsf{s}}$												& 3		&0.647			&0.647		& 0.590			& 0.771			&1.505			& ... 					&25 (1.9\%)		\\
		\noalign{\smallskip}\hline\noalign{\smallskip}
		$\mathsf{c}\bar{\mathsf{u}}/\mathsf{c}\bar{\mathsf{d}}$		& 3		&1.603 			& 0.553 		&0.800			&1.842	 		&2.050 	 	&78  $[5]$			& 30 (1.4\%)	\\
		$\mathsf{c}\bar{\mathsf{s}}$												 & 3	&1.603 			& 0.647			&0.800			&1.944	  		& 2.147	 	&40 $[9]$			& 25 (1.1\%)	\\
		$\mathsf{c}\bar{\mathsf{c}}$												 & 4	&1.603			&1.603			&0.966			&3.017  		&3.139  	&31 $[8]$		 	& 20 (.62\%)	\\
		$\mathsf{b}\bar{\mathsf{u}}/\mathsf{b}\bar{\mathsf{d}}$ 	& 4		&4.902			& 0.553 		&1.067			& 5.291  		& 5.339  	&21 $[4]$			& 6.0 (.11\%)	\\
		$\mathsf{b}\bar{\mathsf{s}}$												&4		&4.902 			&0.647  		& 1.067			& 5.379  		& 5.428 	& 37 $[4]$			& 5.6 (.10\%)	\\
		$\mathsf{b}\bar{\mathsf{c}}$												& 4		&4.902			&1.603			& 1.196		 	& 6.258 	 	& 6.316		&37 $[2]$			& 5.3 (.08\%)	\\
		$\mathsf{b}\bar{\mathsf{b}}$												& 5		&4.902			&4.902			& 1.389			& 9.475 		& 9.514		&38 $[14]$			& 5.6 (.06\%)	\\
		\noalign{\smallskip}\hline
	\end{tabular}
	\caption{Summary of the model parameters with the basis truncation $N_\text{max} = L_\text{max} = 32$. Among them $m_q(m_{\bar{q}})=0.553$ GeV and $0.647$ GeV are the two fitted parameters. The calculated meson masses of the two low-lying states, known as pseudoscalar (PS) and vector (V), are listed in the table. The r.m.s. are the root-mean-square differences of our results from the experimental measured masses; the number of compared states $N_\text{exp}$ are shown in the square brackets. The mean spread $\overline{\delta_J M}$ is the spread in  the masses over allowed $M_J$ values and the deviation from zero reflects the violation of rotational symmetry. Following the absolute mean spread value, we provide the relative spread with respect to the total mass of constituent (anti-) quarks in the parenthesis. 
		Note that the masses of light quarks listed in the table are fitted from heavy-light mesons, and applied to light mesons for supplementary calculations.
		Hence we do not provide the r.m.s. of the light mesons due to the absence of chiral dynamics in this model, which is outside the scope of this work.}
	\label{tb1}
	\vspace*{1cm} 
\end{table*}

\subsection{Mass Spectrum}

\begin{figure*}
	\centering
	\begin{subfigure}{0.45\textwidth}
		\centering
		\includegraphics[scale=0.3]{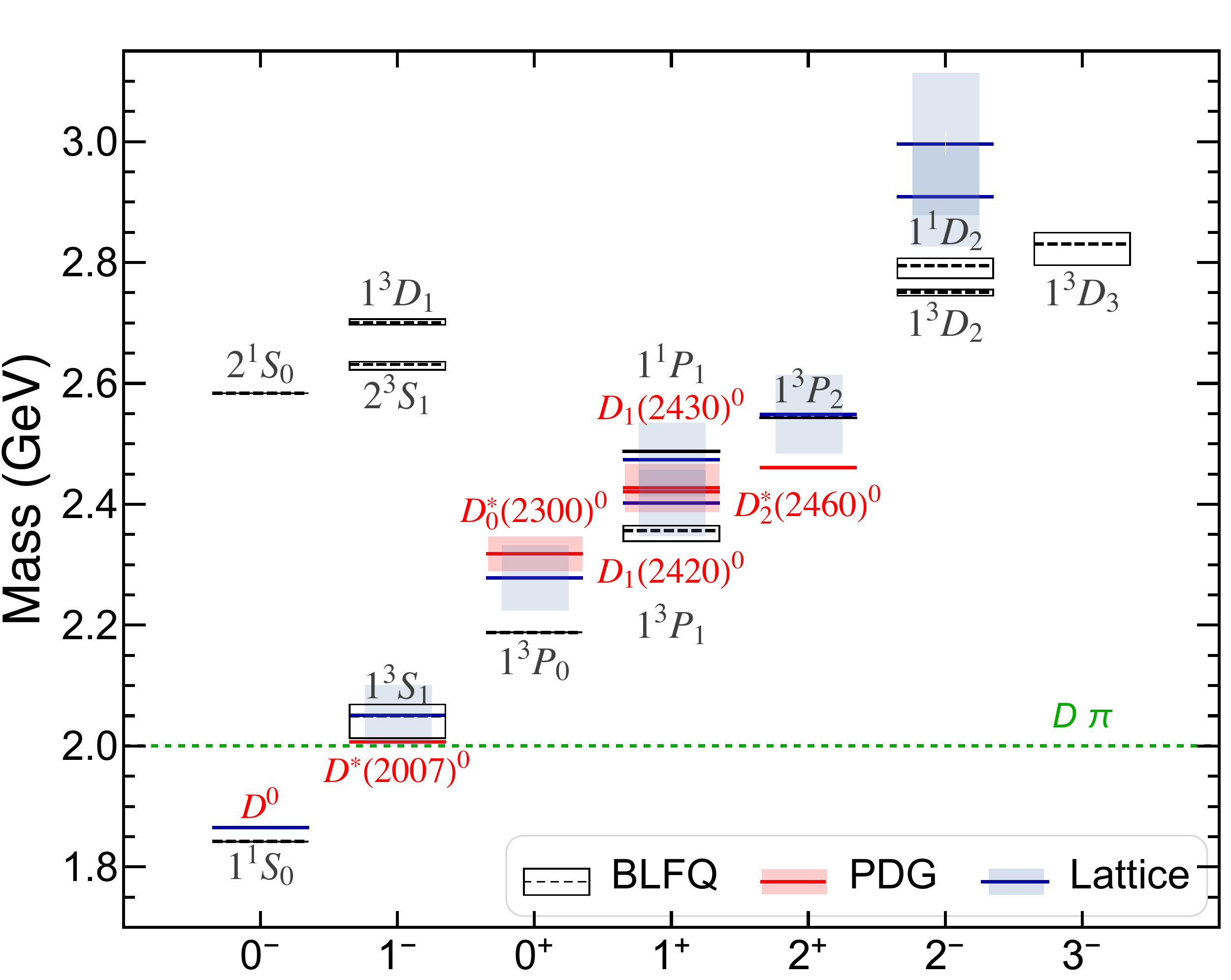} 
		\caption{$D$}
	\end{subfigure}
	\hspace{0.5cm}
	\begin{subfigure}{0.45\textwidth}
		\centering
		\includegraphics[scale=0.3]{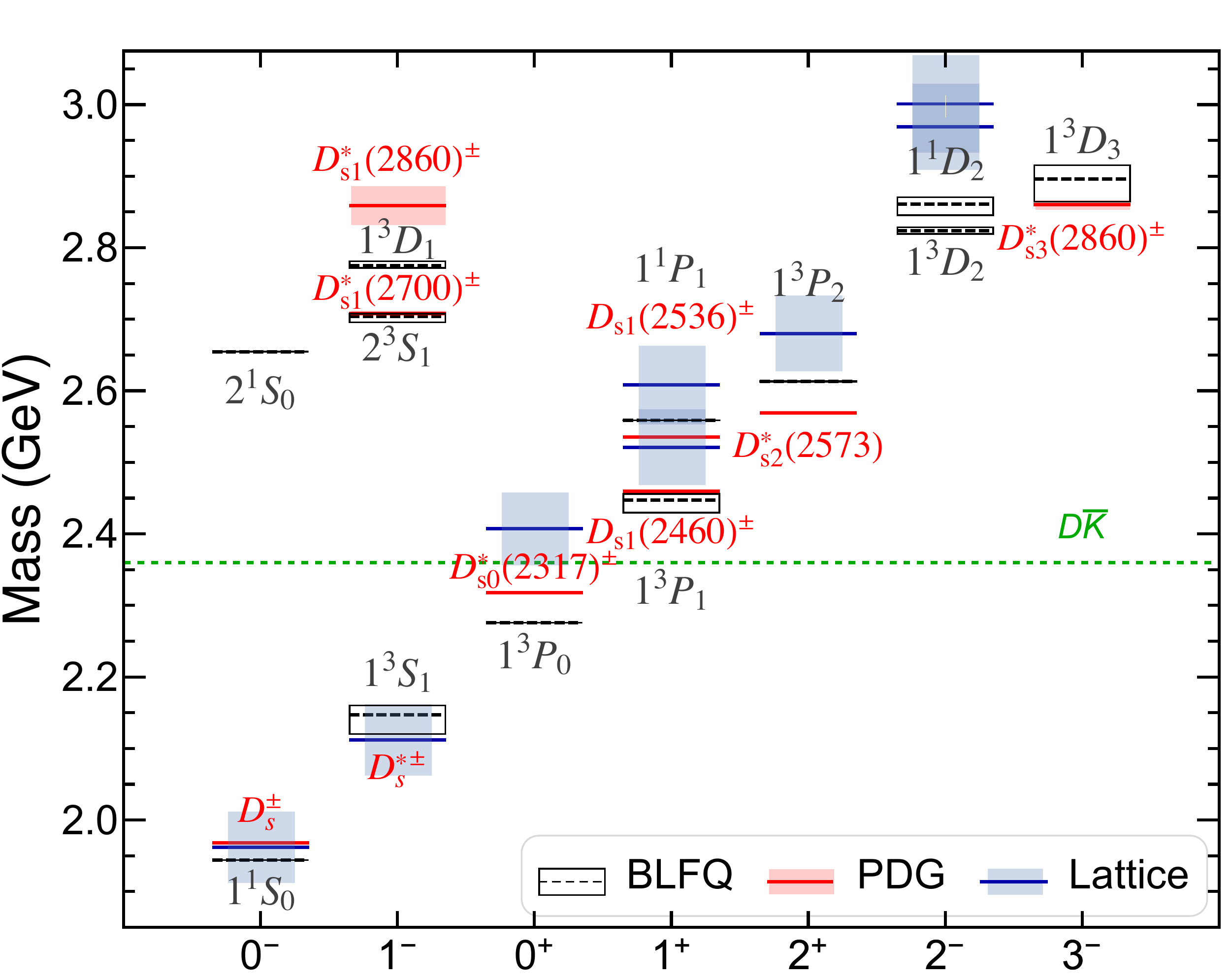}
		\caption{$D_s$}
	\end{subfigure}
	\\
	\begin{subfigure}{0.45\textwidth}
		\centering
		\includegraphics[scale=0.3]{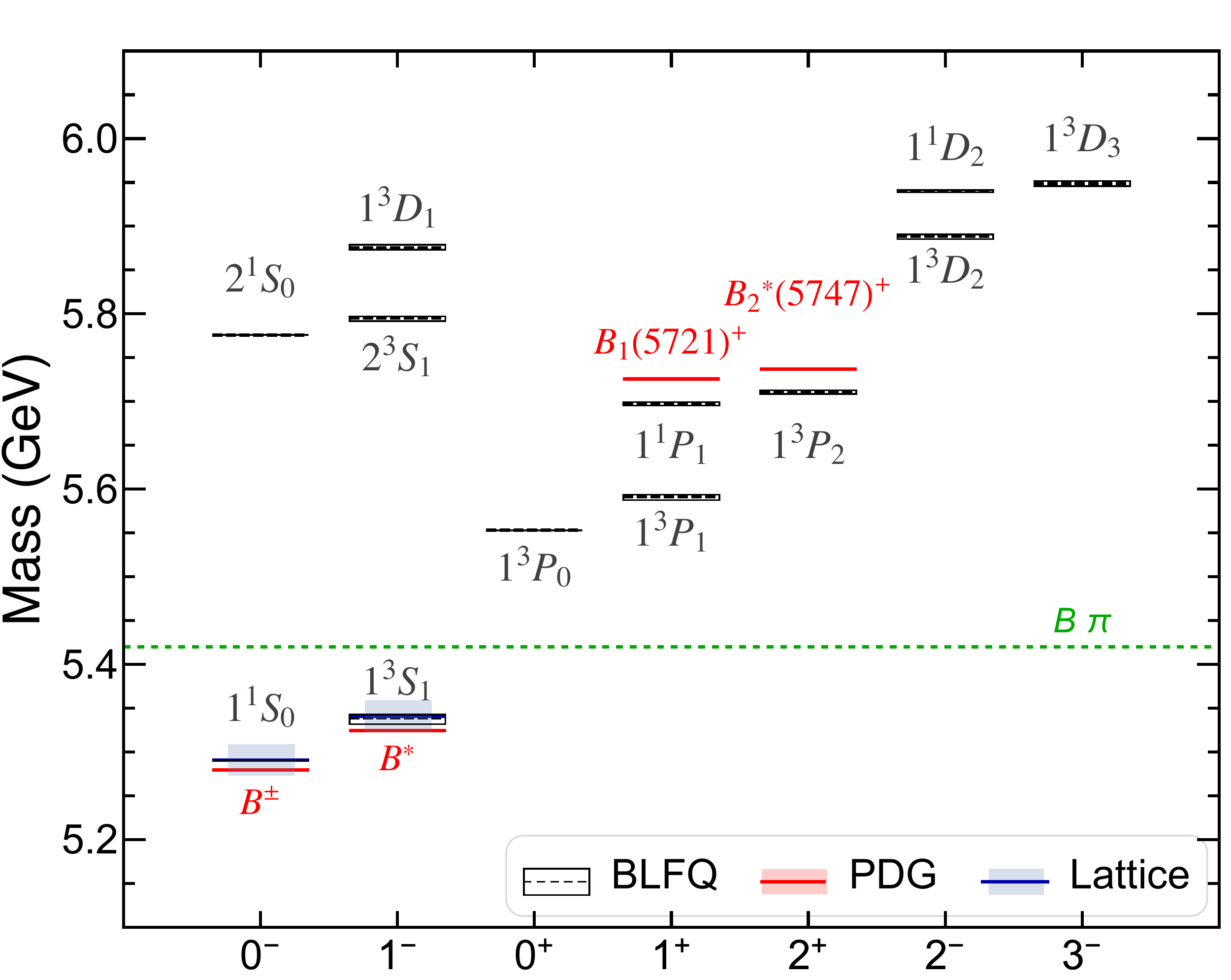}
		\caption{$B$}
	\end{subfigure}
	\hspace{0.5cm}
	\begin{subfigure}{0.45\textwidth}
		\centering
		\includegraphics[scale=0.3]{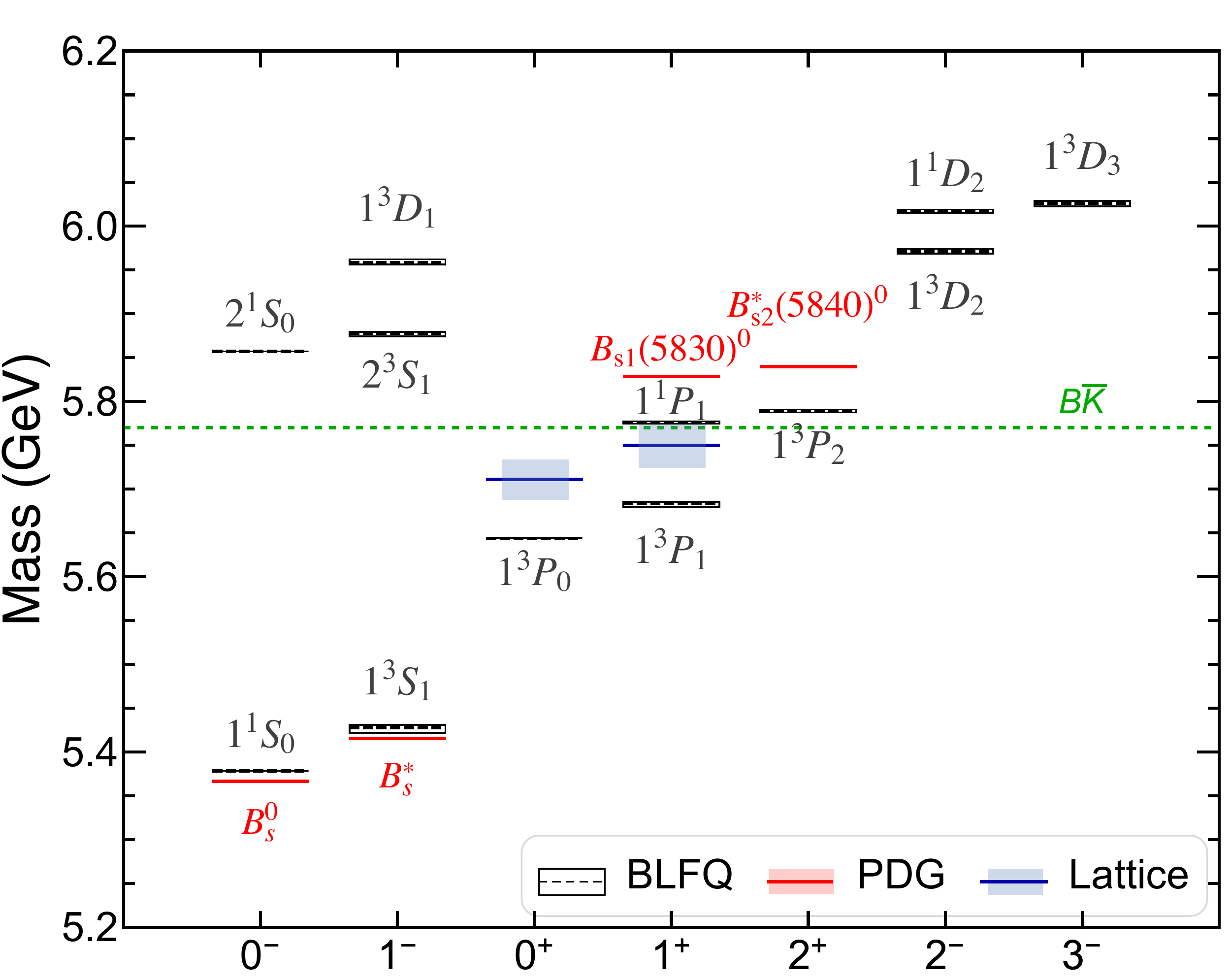}
		\caption{$B_s$}
	\end{subfigure}
	
	\caption{	\label{fig1}Mass spectra of the four heavy-light meson systems, each with 12 low-lying states, using $N_\text{max}=L_\text{max}=32$. Horizontal axis refers to the $J^P$ values of the states. The green dashed lines indicate the corresponding open flavor threshold. Black boxes are the results of this work, indicating the highest and lowest mass from different $m_J$'s, while the dashed lines in between are the averages $\overline{M}$. Red bars are the experimental masses compiled by PDG~\cite{PhysRevD.98.030001}. Specifically, we use the values of $D^0 \ (\mathsf{c}\bar{\mathsf{u}})$ and $B^\pm \ (\mathsf{b\bar{\mathsf{u}}}/\mathsf{b}\bar{\mathsf{d}})$ from experiments for comparison. Shaded blue bars are the Lattice QCD results from Refs~\cite{PhysRevD.83.014506,PhysRevD.86.094510,LANG201517,PhysRevD.92.094508}. For both PDG and Lattice, we use the solid lines for the center values, while shaded boxes to indicate the uncertainties.}
	\vspace*{1cm}
\end{figure*}

Here we present mass spectra of the four heavy-light meson systems, with all 12 low-lying states in Fig.~\ref{fig1}. 
For each of the BLFQ calculated states, we show the mass as a black box with a dashed line, where the dashed line represents the mean value of the invariant mass,
\begin{equation}
\overline{M} = \sqrt{\frac{M^2_{-J}+M^2_{-J+1}+...+M^2_{J}}{2J+1}}\ ;
\end{equation}
and the box indicates the spread of eigenmasses obtained from different $m_J$'s:  $\delta_J M \equiv \max(M_{m_J}) -  \min(M_{m_J})=$ box height.
In many cases, the box height is small and not visible in the figure. Since rotational symmetry would imply degeneracy (zero box height), we introduce an overall mean spread for hadrons ($h$) within a fixed flavor to characterize the violation of rotational symmetry,
\begin{equation}
\overline{\delta_J M} \equiv \sqrt{\frac{1}{N_h}\sum_{h}^{J \ne 0} (\delta_J M_h)^2} \qquad  \Big(N_h \equiv \sum_{h}^{J \ne 0} 1 =9\Big),
\end{equation} 
listed in the last column of Table~\ref{tb1}. For each system, $\overline{\delta_J M}$ is obtained with 9 states whose $J > 0$, which includes three $ J^P = 1^-$ states, two $1^+$, one $2^+$, two $2^-$, and one $3^-$. We observe that the mean spread has the tendency of decreasing with increasing meson mass, which is in agreement with our expectation.  Viewed as a percentage deviation (see the parentheses in the last column of Table~\ref{tb1}), the decrease with increasing meson mass is in a trend consistent with the trend found for all light and heavy mesons~\cite{PhysRevD.96.016022,PhysRevD.98.114038}.

%	The overall layouts of these states for the four systems are comparable. 

The hyperfine splitting of ground state pseudoscalar and vector is a particularly sensitive test of the spin-sensitive component of the Hamiltonian -- the effective one-gluon exchange interaction. 
It is well-known that for the light mesons, this mass splitting is driven by the chiral symmetry breaking~\cite{PhysRevC.56.3369}.
However, in our approach, the splitting is brought in by the one-gluon exchange interaction and its interplay with the confining strength $\kappa$ when the basis size is fixed by $N_\text{max}$ and $L_\text{max}$.
Specifically, smaller $\kappa$ tends to lead a larger mass splitting between the two states. The size of the hyperfine splittings are reasonable for heavy mesons~\cite{PhysRevD.96.016022,PhysRevD.98.114038,LI2016118} and heavy-light mesons within this model. However, we observe that for BLFQ to reproduce the properties of the light mesons, such as the Goldstone nature of the pions in the chiral limit, the Hamiltonian needs to include chiral dynamics~\cite{PhysRevC.99.035206}.

%	We have tested our model with another confining strength $\kappa = 0.54$ GeV provided in Ref. \cite{BRODSKY20151}, and found the overall spectra were not significantly affected: the absolute variation is no more than 5 MeV in either r.m.s or $\overline{\delta_J M} $ through the four heavy-light systems presented in Table~\ref{tb1}.

We compare our mass spectra with experimental values summarized by the Particle Data Group (PDG)~\cite{PhysRevD.98.030001}, and Lattice QCD~\cite{PhysRevD.83.014506,PhysRevD.86.094510,LANG201517,PhysRevD.92.094508} in Fig.~\ref{fig1}.
Most of our results are within the quoted uncertainties of either experiments or Lattice. Some states, $J^P = 0^+$ or $1^+$ for instance, show somewhat larger differences among the states compared. In BLFQ, the accuracy may be enhanced by introducing higher Fock sectors, such as including a dynamical gluon in the Fock sectors. 
Nevertheless, our work provides the LFWFs that can be used to calculate some hadron observables, which may be tested by ongoing and forthcoming experiments, such as the Belle \textrm{II} experiment at the SuperKEKB accelerator~\cite{kou2018belle}.

\subsection{Light-Front Wave Function}

\begin{figure*}
	\centering
	\begin{subfigure}{0.45\textwidth}
		\centering
		\includegraphics[scale=0.32]{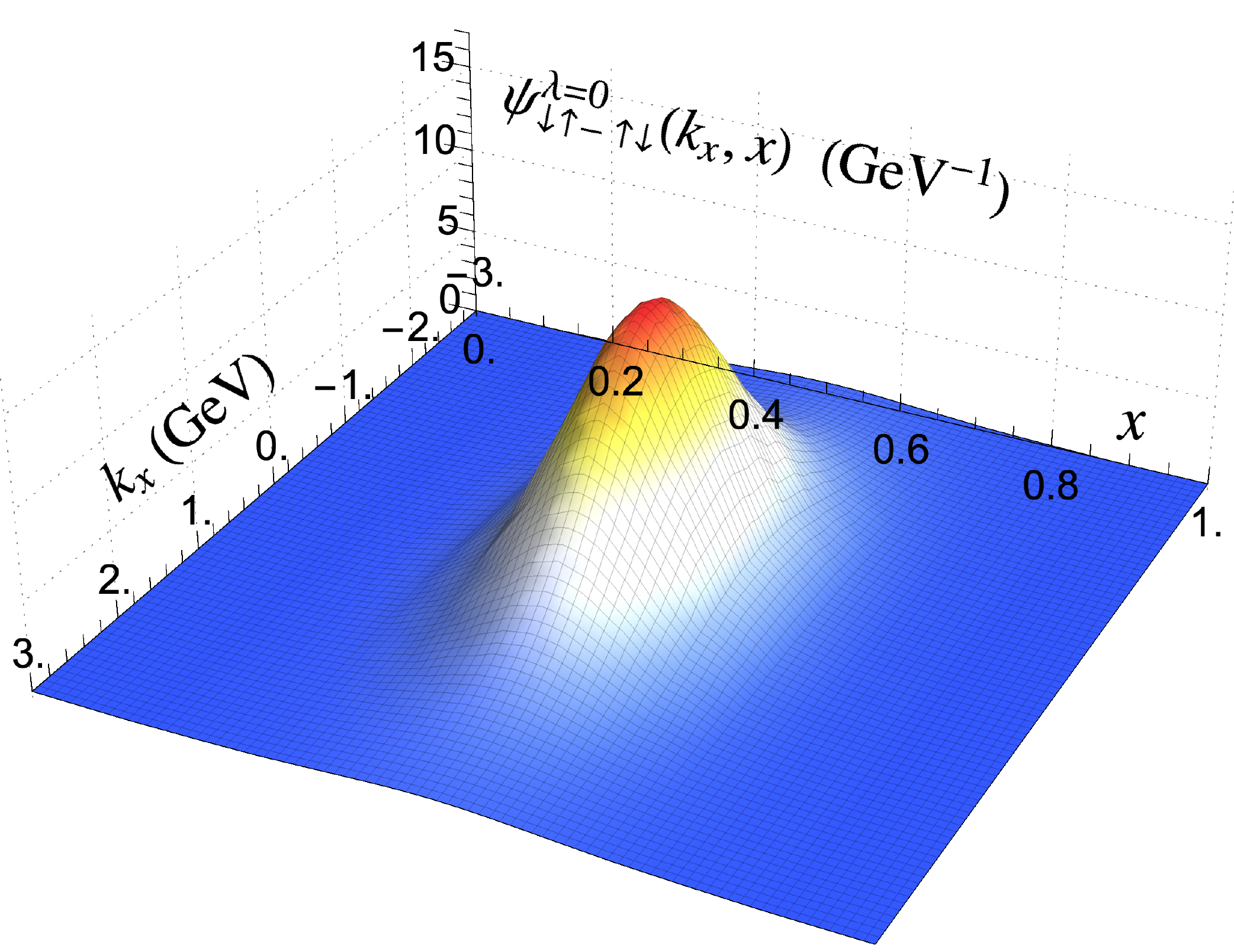} 
		\caption{$\eta_b\ (\mathsf{b}\bar{\mathsf{b}})$\label{fig5a}}
	\end{subfigure}
	\hspace{1cm}
	\begin{subfigure}{0.45\textwidth}
		\centering
		\includegraphics[scale=0.32]{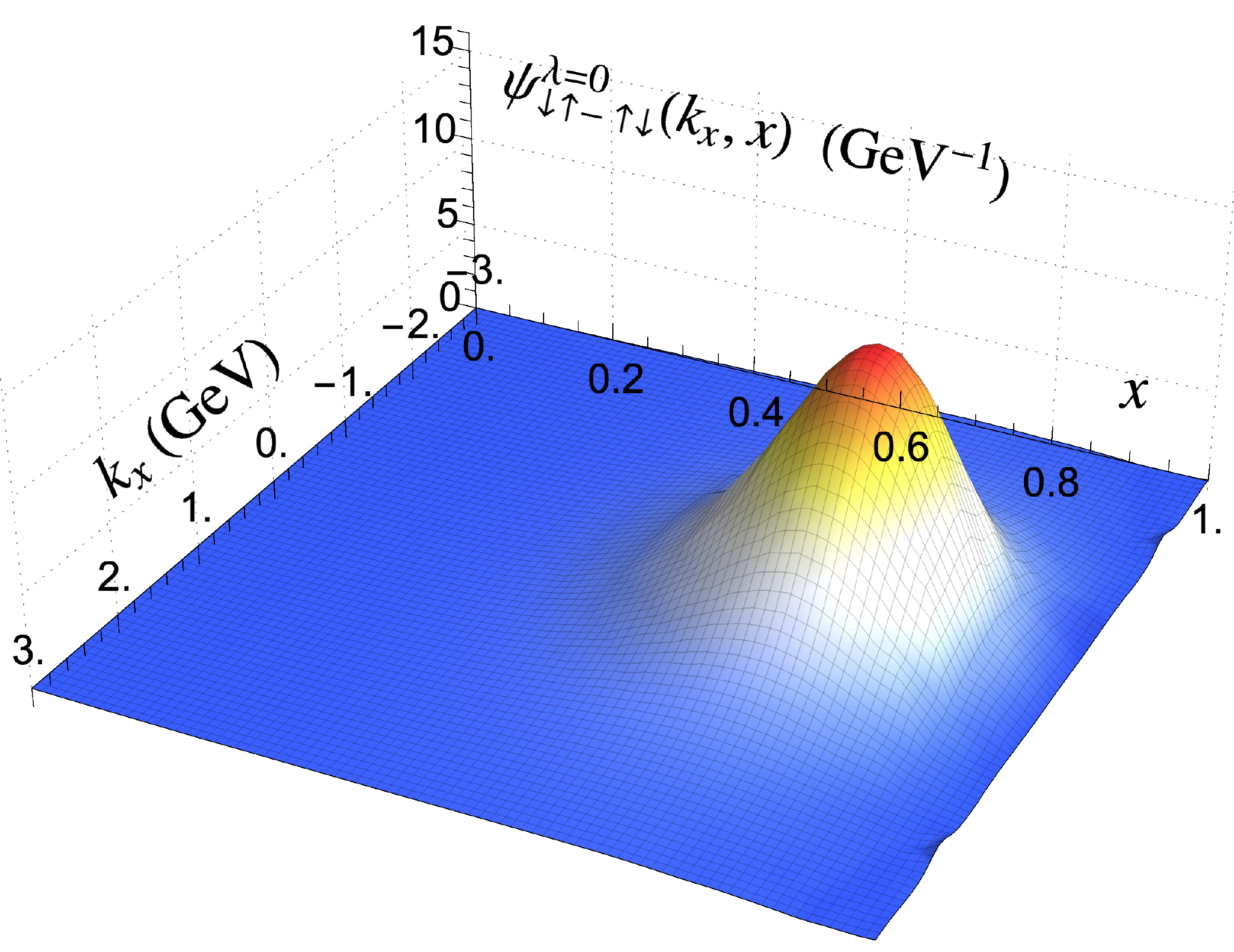}
		\caption{$B_c \ (\mathsf{b}\bar{\mathsf{c}})$\label{fig5b}}
	\end{subfigure} \\ 
	\begin{subfigure}{0.45\textwidth}
		\centering
		\includegraphics[scale=0.32]{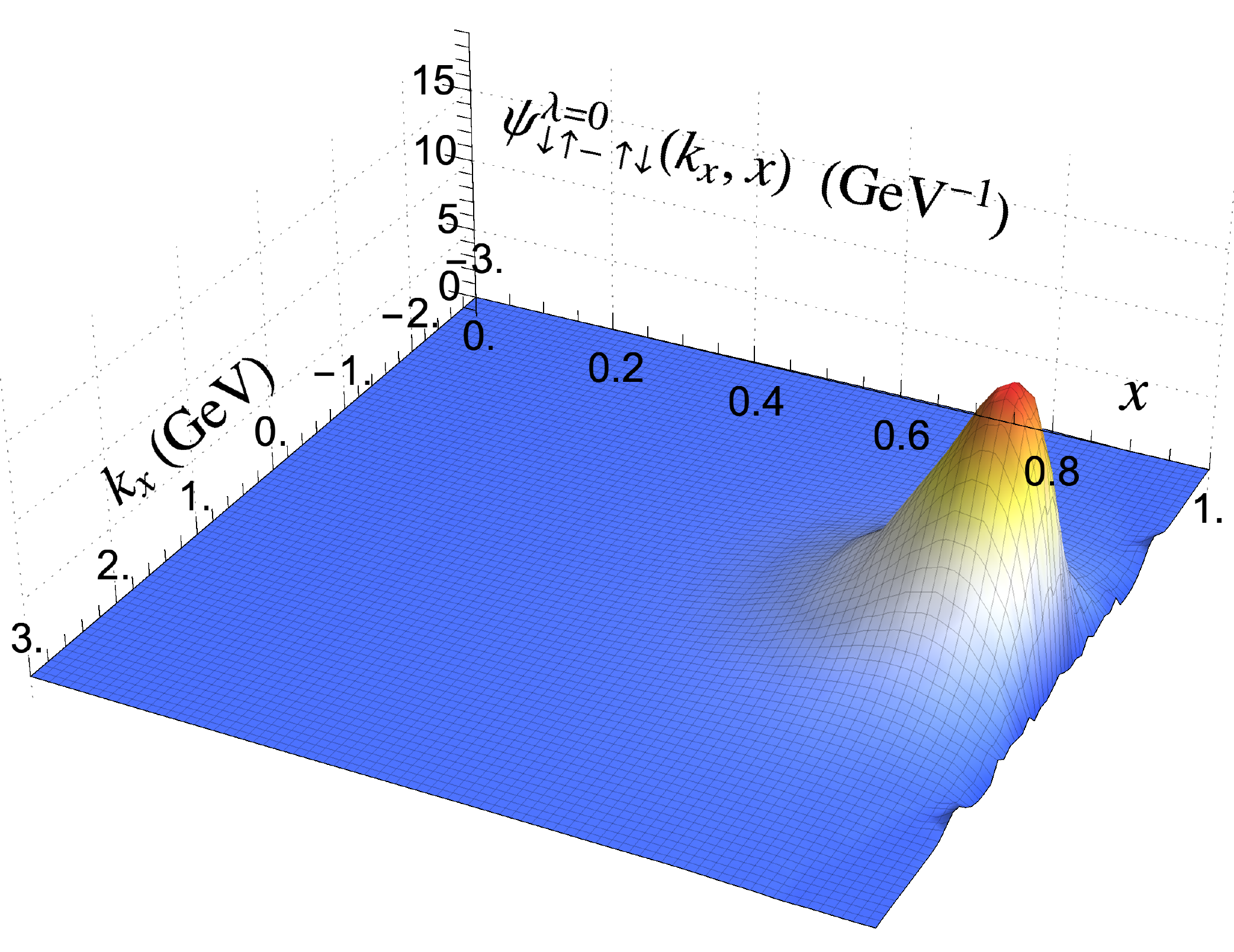}
		\caption{$B \ (\mathsf{b}\bar{\mathsf{u}})$\label{fig5c}}
	\end{subfigure}
	\caption{\label{fig5}The ground state ($1^1S_0$) LFWFs of $\eta_b$, $B_c$, and $B$ at $\theta_q=0$. Locations of peaks along $x$-axis vary significantly according to the quark mass fractions of each system.}
	\vspace*{1cm}	
\end{figure*}

The LFWFs obtained from solving the light-front Hamiltonian matrix eigenvalue problem provide direct access to hadron structure observables. The orthonormality condition of the LFWFs reads,
\begin{equation}
\label{eq2}
\begin{aligned}
&\sum_{s\bar{s}}\int_0^1 \frac{\dd x}{2x(1-x)} \int \frac{\dd ^2 k_\perp}{(2\pi)^3} 	\psi_{s\bar{s}/h'}^{(m'_J)*} (x, \vec{k}_\perp)  	\psi_{s\bar{s}/h}^{(m_J)} (x, \vec{k}_\perp)  \\
&= \delta_{hh'} \delta_{m_J,m'_J}.
\end{aligned}
\end{equation} 
In fact, the LFWFs themselves, especially the unequal mass systems, illustrate interesting features. The equal-mass quarkonium LFWFs exhibit symmetry with respect to the transverse momentum and longitudinal momentum fraction. We anticipate and observe that asymmetry emerges when the constituent masses differ. In Fig.~\ref{fig5}, we present the ground state spin-singlet LFWFs 
$\psi^{m_J=0}_{\uparrow\downarrow-\downarrow\uparrow}(\vec{k}_\perp,x) $ 
of three meson states, $\eta_b(\mathsf{b}\bar{\mathsf{b}})$, $B_c(\mathsf{b}\bar{\mathsf{c}})$, and $B(\mathsf{b}\bar{\mathsf{u}})$, where we define 
$\psi^{m_J=0}_{\uparrow\downarrow\pm\downarrow\uparrow}(\vec{k}_\perp,x) \equiv 1/\sqrt{2}[ \psi^{m_J=0}_{\uparrow\downarrow}(\vec{k}_\perp,x) \pm \psi^{m_J=0}_{\downarrow\uparrow}(\vec{k}_\perp,x) ] $. For the sake of convenience, we visualize the LFWFs at $k_y=0$, i.e. $\theta_q=0$ or $\pi$ in the phase $\exp(\imag m \theta_q)$ (cf. Eq.~\eqref{eq4}).  

\begin{figure*}
	\centering
	\begin{subfigure}{0.45\textwidth}
		\centering
		\includegraphics[scale=0.32]{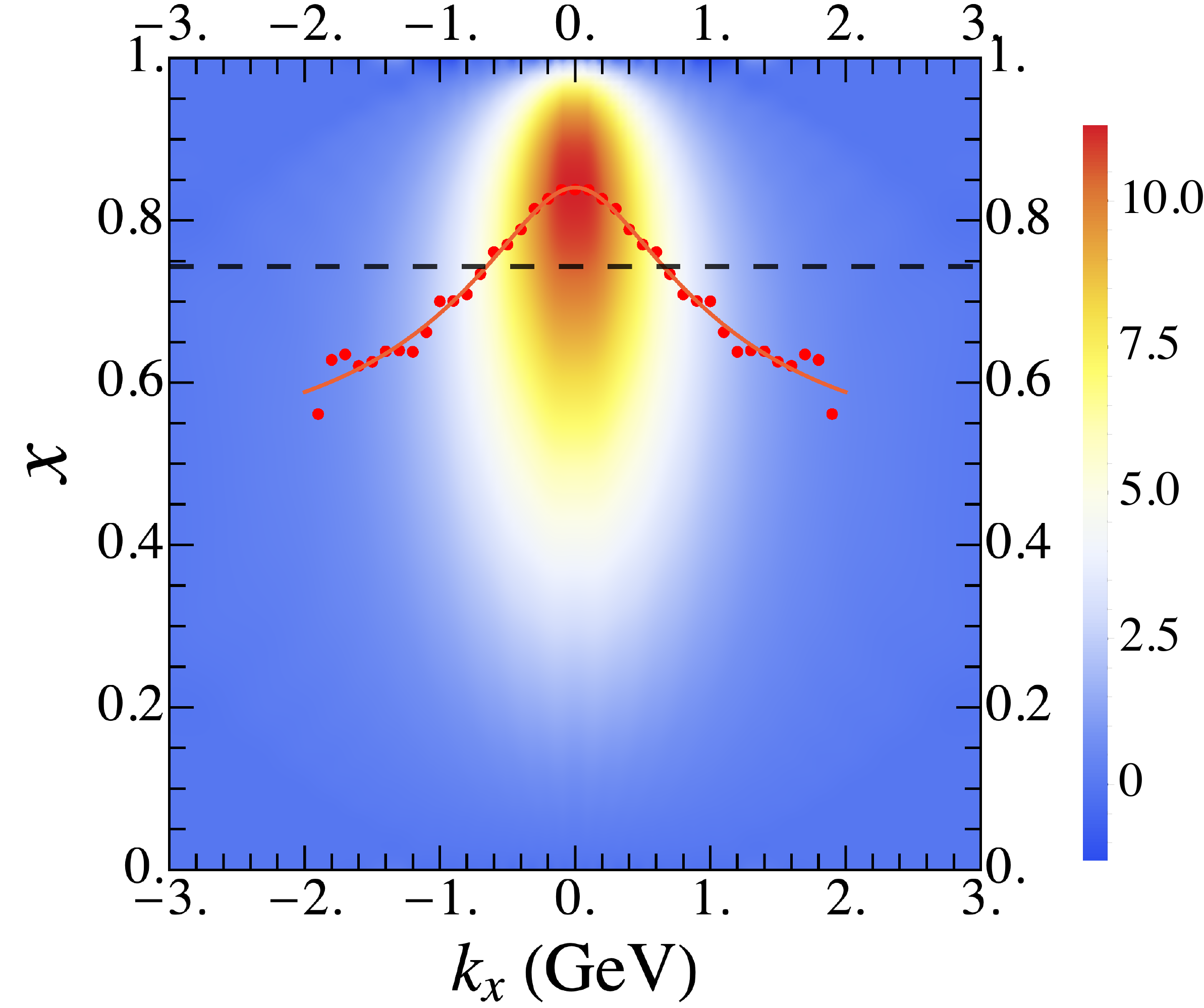} 
		\caption{$D\ (\mathsf{c}\bar{\mathsf{u}})$\label{fig6a}}
	\end{subfigure}
	\hspace{1cm}
	\begin{subfigure}{0.45\textwidth}
		\centering
		\includegraphics[scale=0.32]{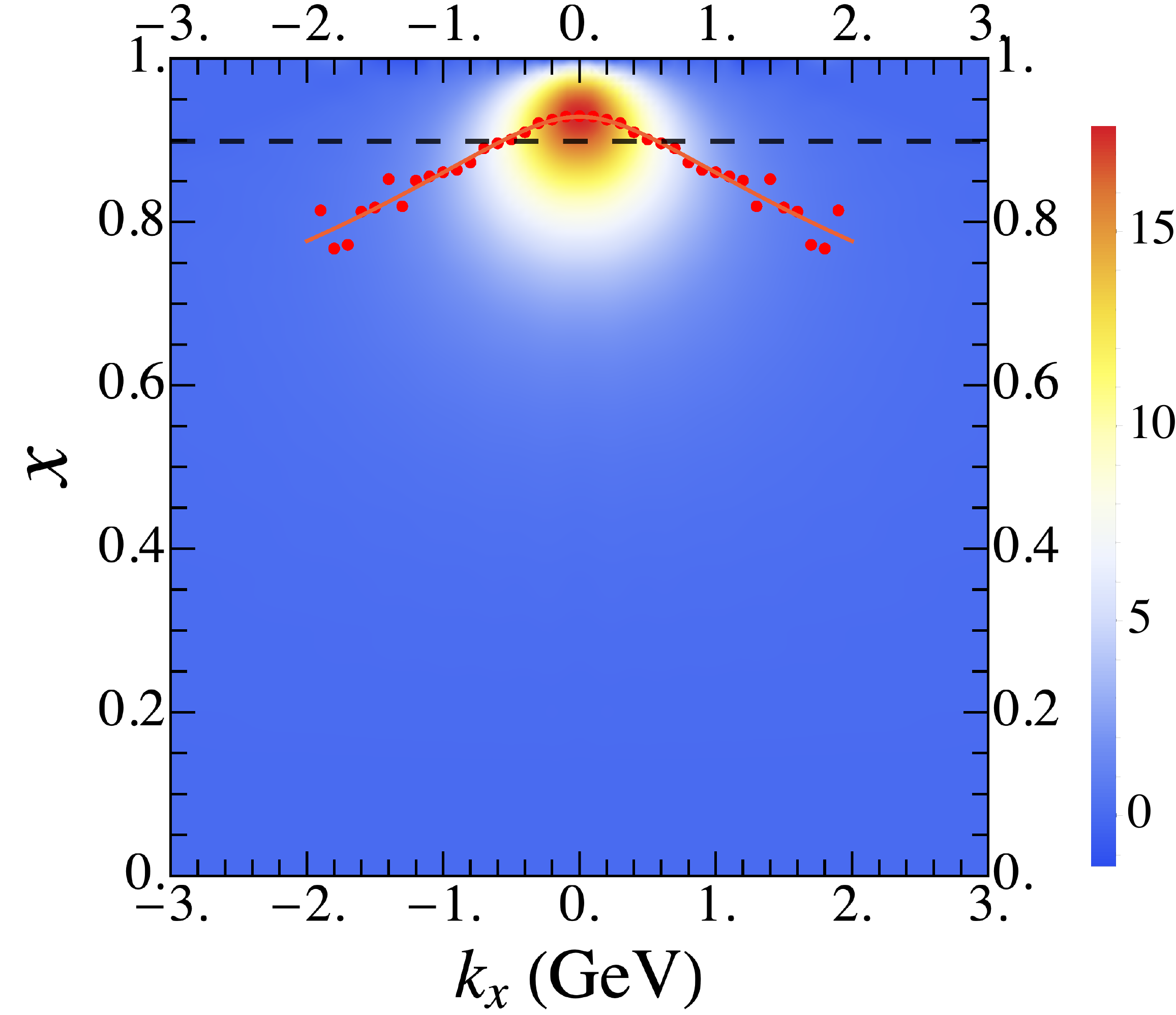}
		\caption{$B \ (\mathsf{b}\bar{\mathsf{u}})$\label{fig6b}}
	\end{subfigure}
	\caption{\label{fig6}Density plot of the ground state ($1^1S_0$) LFWFs of $D$ and $B$ mesons.  The dashed black line represents the quark mass fraction, which is $x=m_\mathsf{c(b)}/(m_\mathsf{c(b)}+m_\mathsf{u})$ for $D$ ($B$) meson. Red dots indicate the local maxima along $x$ for fixed $k_\perp$, and the orange curve is fitted from the dots based on the light-front parton gas model~\cite{Jia:2018hxd,Maris:2020wew}, to illustrate the trend of peaks: for the heavy-light mesons, the peak is located at $x>m_q/(m_q+m_{\bar{q}})$ when $k_\perp=0$, and it shifts to smaller $x$ as ${k_\perp}$ increases and finally approaches to $x=1/2$ at the limit ${k_\perp} \rightarrow \infty$.}
	\vspace*{1cm}
\end{figure*}

The LFWFs of three listed mesons are symmetric in the transverse direction.  The equal-mass (quarkonium) case, Fig.~\ref{fig5a}, also shows the expected symmetry in the longitudinal direction.  However, the unequal-mass cases, Figs. \ref{fig5b} and \ref{fig5c}, show the anticipated asymmetry in the longitudinal direction.
Specifically, the peak of LFWFs for the quarkonia is located at $x=1/2$ on the longitudinal direction, which is the same as the quark mass fraction $m_q/(m_q+m_{\bar{q}})$ where $m_q = m_{\bar{q}}$. 
However, for the unequal-mass system, the peak location is nontrivial as shown in Fig.~\ref{fig6}, where we present the density plot of the ground state spin-singlet LFWFs of $D$ and $B$. 
%the peak along $x$-axis tends to decrease with $k_x$, and ultimately approach to $x=1/2$ when $k_x \rightarrow \infty$.
%The presence or absence of symmetries affects other observables, such as the parton distribution function as we will discuss later in Sec.~\ref{subs3}.
At $k_\perp = 0$ the asymmetry is maximal, and the peak is located at $x > m_\mathsf{c(b)} / (m_\mathsf{c(b)} + m_\mathsf{u})$, but as $k_\perp$ increases the peak location (along fixed $k_\perp$) shifts to smaller values of $x$.  In the limit $k_\perp \rightarrow \infty$ the peak location approaches $x = 1/2$.  This can easily be understood because for $k_\perp \gg m_\mathsf{c(b)}+m_\mathsf{u}$, the quark masses become irrelevant, and the LFWF approaches that of equal-mass constituents. These features confirm the prediction from parton gas model~\cite{Jia:2018hxd}. The nontrivial dependence of the LFWF on the quark masses can also be found in $B_c$ meson~\cite{Maris:2020wew}, and it affects light-front observables such as the parton distribution function (PDF) and distribution amplitude (DA) as we will see in the following sections.

\subsection{Parton Distribution Function \label{subs3}}

The structure of hadrons, as probed in inclusive deep inelastic scattering (DIS) at large momentum transfer, can be described by the PDF $f(x; \mu)$. The PDF within collinear factorization defines the probability for finding a quark carrying a longitudinal momentum fraction $x$ at resolution scale $\mu$. In the LFWF representation, the PDF is calculated simply by integrating out the transverse momentum of the square of the wave function modulus,
\begin{equation}
f(x;\mu) =\frac{1}{4\pi x(1-x)} \sum_{s\bar{s}} \int^{\lesssim \mu^2} \frac{\dd ^2 k_\perp}{(2\pi)^2} \abs{\psi_{s\bar{s}} (x, \vec{k}_\perp)}^2,
\end{equation}
and the truncation parameter $N_\text{max}$ of the LFWFs provides a UV scale for model $\mu \approx \kappa \sqrt{N_\text{max}}$, which is denoted schematically by the upper limit in the integration. That is, the integration is performed without any additional cutoff.
Within the two-body Fock sector truncation, the PDF is normalized to unity $\int_0^1 f(x; \mu) \dd x  =1$, 
which relates to the orthonormal condition of the LFWFs (cf. Eq.~\eqref{eq2}).

\begin{figure*}
	\centering
	\begin{subfigure}{0.45\textwidth}
		\centering
		\includegraphics[scale=0.35]{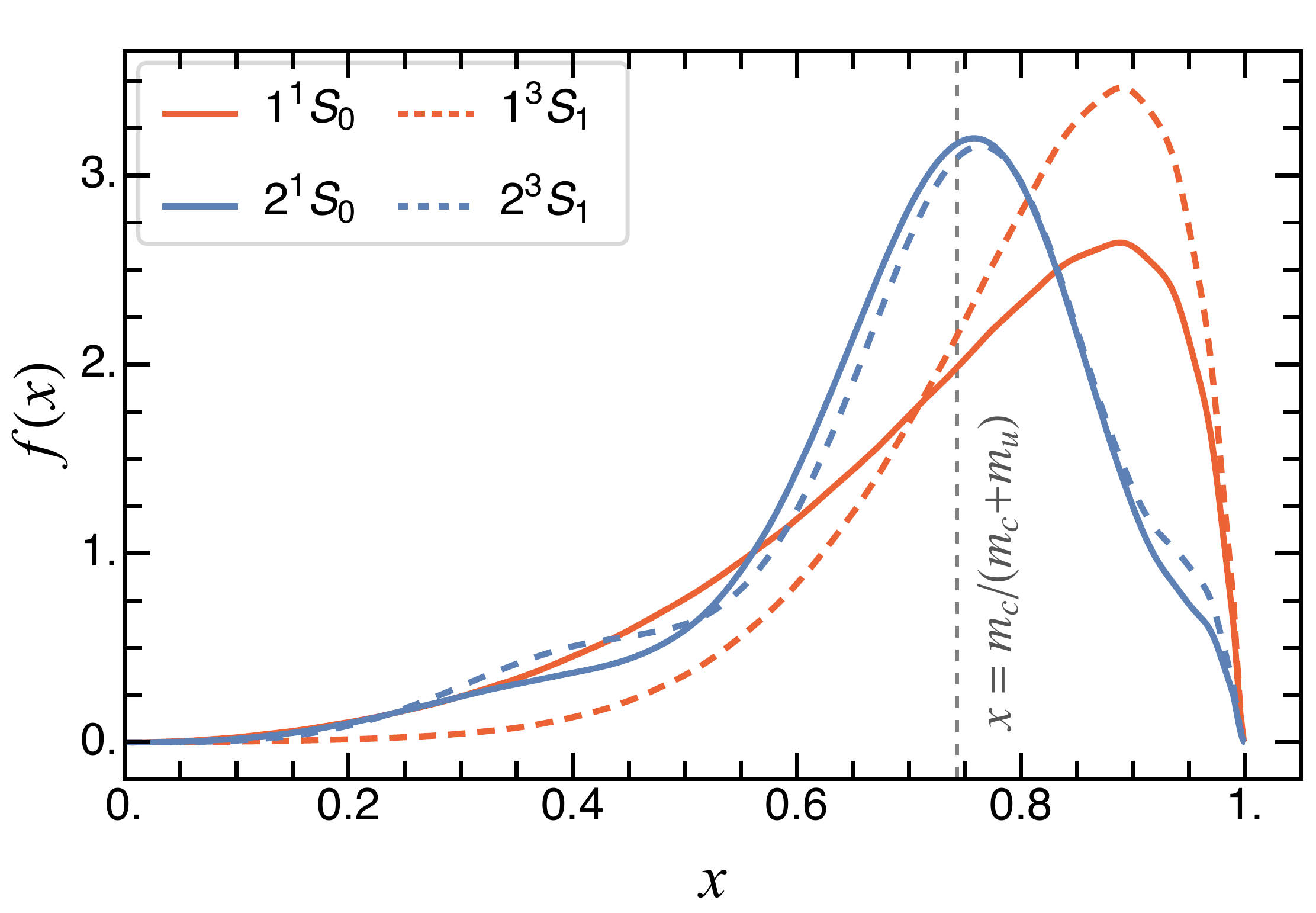} 
		\caption{$D$}
	\end{subfigure}
	\hspace{.5cm}
	\begin{subfigure}{0.45\textwidth}
		\centering
		\includegraphics[scale=0.35]{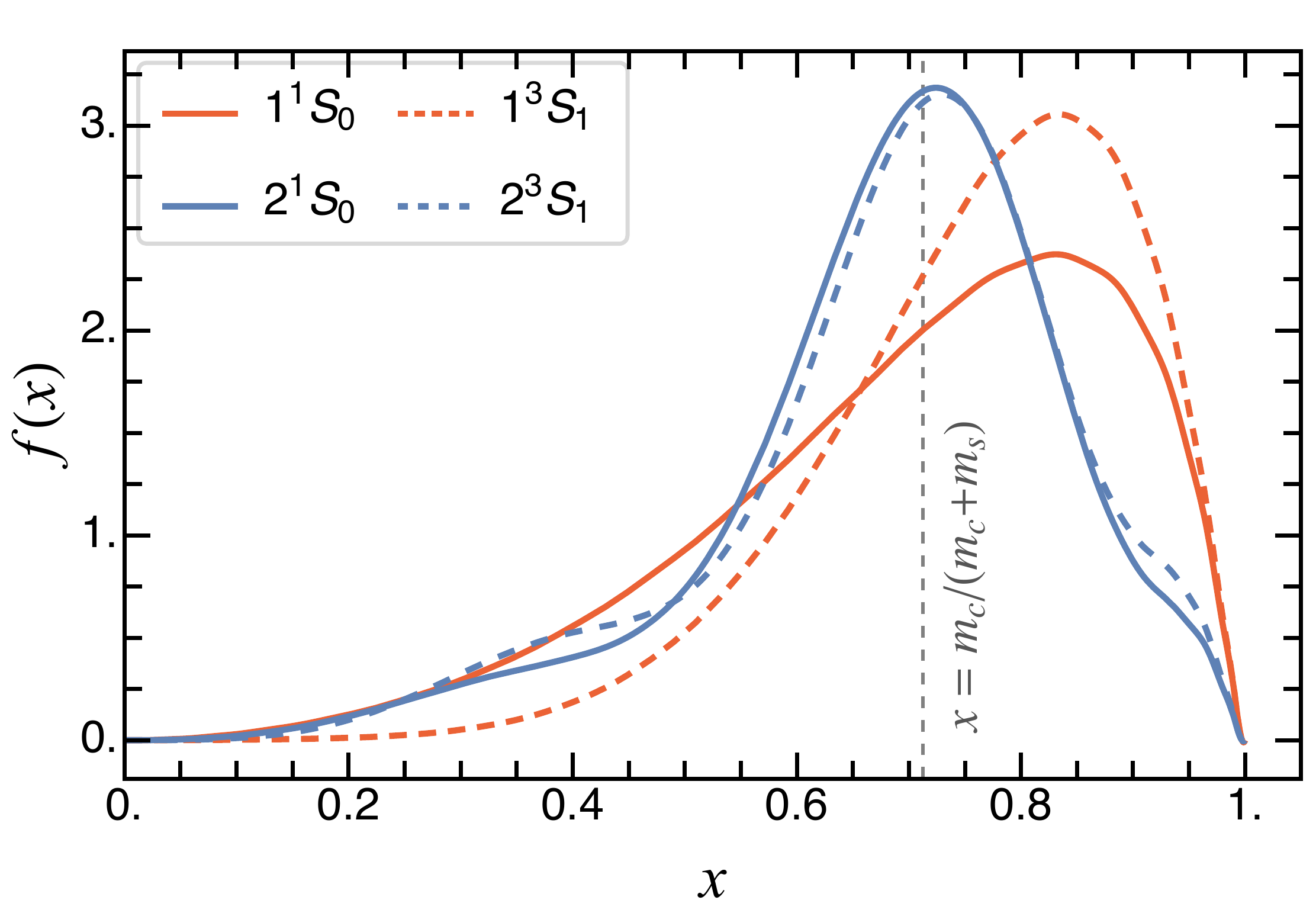}
		\caption{$D_s$}
	\end{subfigure} \\
	\begin{subfigure}{0.45\textwidth}
		\centering
		\includegraphics[scale=0.35]{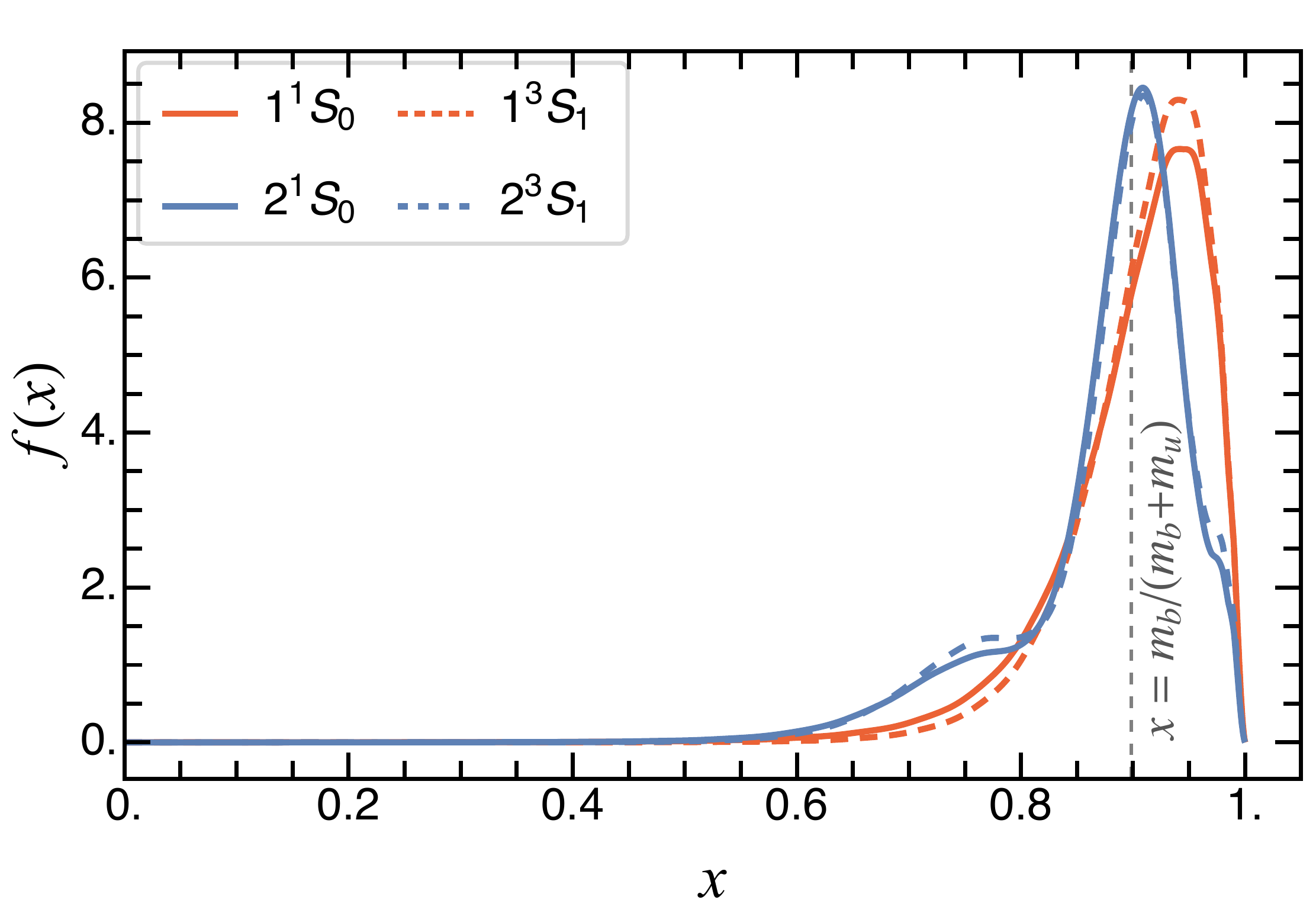} 
		\caption{$B$}
	\end{subfigure}
	\hspace{.5cm}
	\begin{subfigure}{0.45\textwidth}
		\centering
		\includegraphics[scale=0.35]{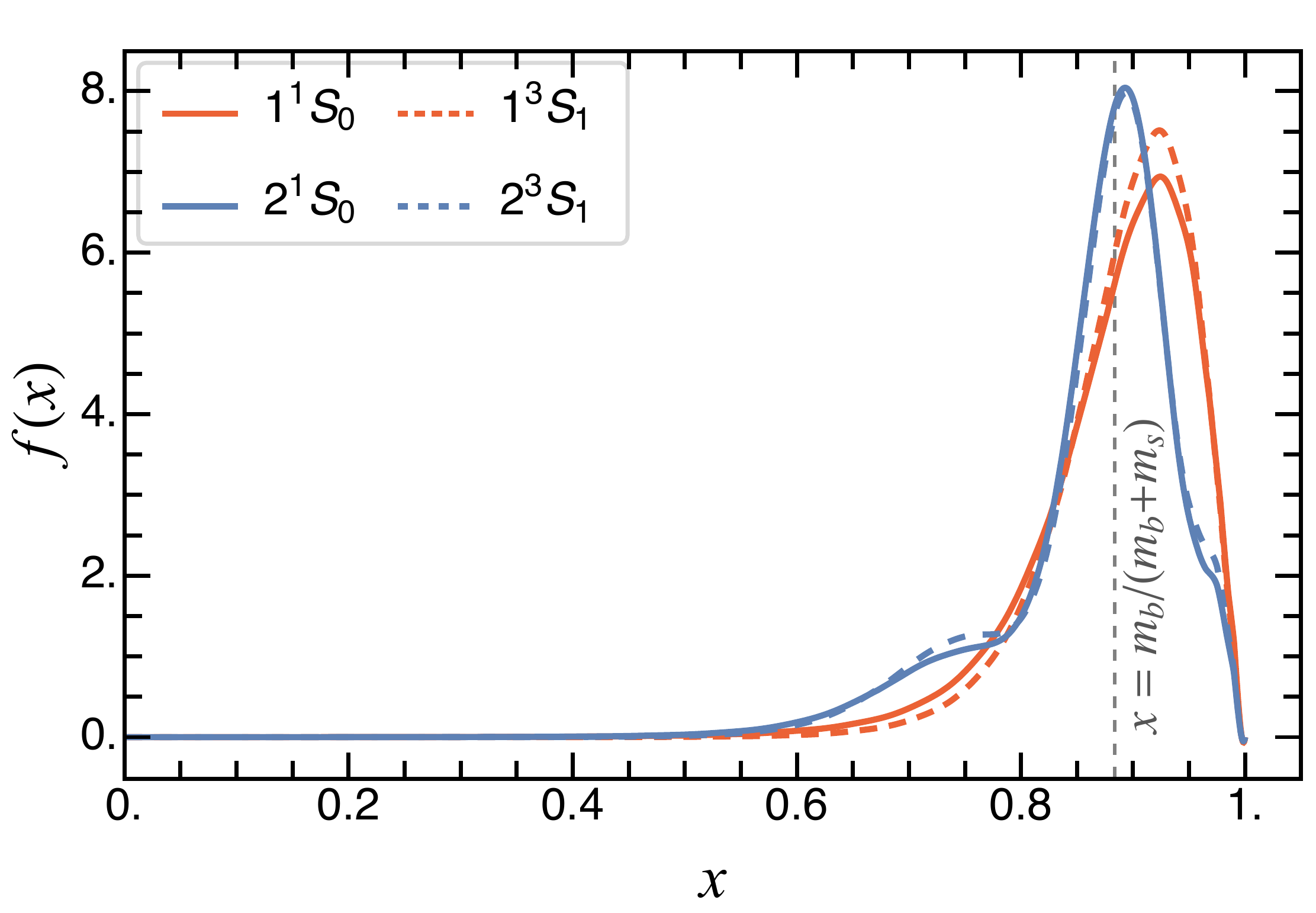}
		\caption{$B_s$}
	\end{subfigure} \\	
	\caption{\label{fig4} PDFs of pseudoscalar (solid curves) and vector (dashed curves) states of the heavy-light systems at $N_\text{max} = L_\text{max} = 32$, which is equivalent to the UV regulators $\mu_{D/D_s} \approx4.5$ GeV and $\mu_{B/B_s}\approx6.0$ GeV. The vertical dashed gray lines indicate the mass fraction of the quark, $x=m_q/(m_q+m_{\bar{q}})$.}
	\vspace*{1cm}
\end{figure*}

We study the PDFs of the heavy-light systems, and present the results of pseudoscalar and vector states in Fig.~\ref{fig4}. The location of peak reflects the asymmetry of the quark mass distribution in the heavy-light systems. Note that the peak of the PDFs is not at the quark mass fraction, due to the nontrivial behavior of the maxima in $x$ along fixed $k_\perp$ in the LFWFs. We find that the peak in the PDF for the radial excited states $2^1S_0$ and $2^3S_1$ is at a significantly smaller value of $x$ than that of  $1^1S_0$ and $1^3S_1$, and in fact it is very close to $x = m_q / (m_q+m_{\bar{q}})$. This latter may be coincidental.
The width of PDFs decreases as the mass of the system increases, and approaches a $\delta$ function in the nonrelativistic limit. The $2^1S_0$ and $2^3S_1$ states contain bumps on both sides of the peaks incorporating features arising from radial excitations. The difference between solid and dashed curves reveals the spin excitation. Those differences between $1^1S_0$ and $1^3S_1$ states are larger than between $2^1S_0$ and $2^3S_1$, that is due to the significant hyperfine splittings in lower excited states.

Note that we use the basis size as $N_\text{max} = L_\text{max} = 32$, which corresponds to different resolution scales: $\mu_{D/D_s} \approx4.5$ GeV and $\mu_{B/B_s}\approx6.0$ GeV. 
The PDFs obtained here are only for valence partons with a low resolution where the valence Fock sector approximation is reasonable. Ref.~\cite{PhysRevLett.122.172001} shows that combining the low resolution PDFs with the  Dokshitzer-Gribov-Lipatov-Altarelli-Parisi (DGLAP) evolution provides access to experiment-relevant PDFs (valence, sea and gluon) at higher scales.  We anticipate that DGLAP could also be applied to our results but that is beyond the scope of the present effort.

\subsection{Distribution Amplitude \label{subs4}}
\begin{figure*}
	\centering
	\begin{subfigure}{0.45\textwidth}
		\centering
		\includegraphics[scale=0.35]{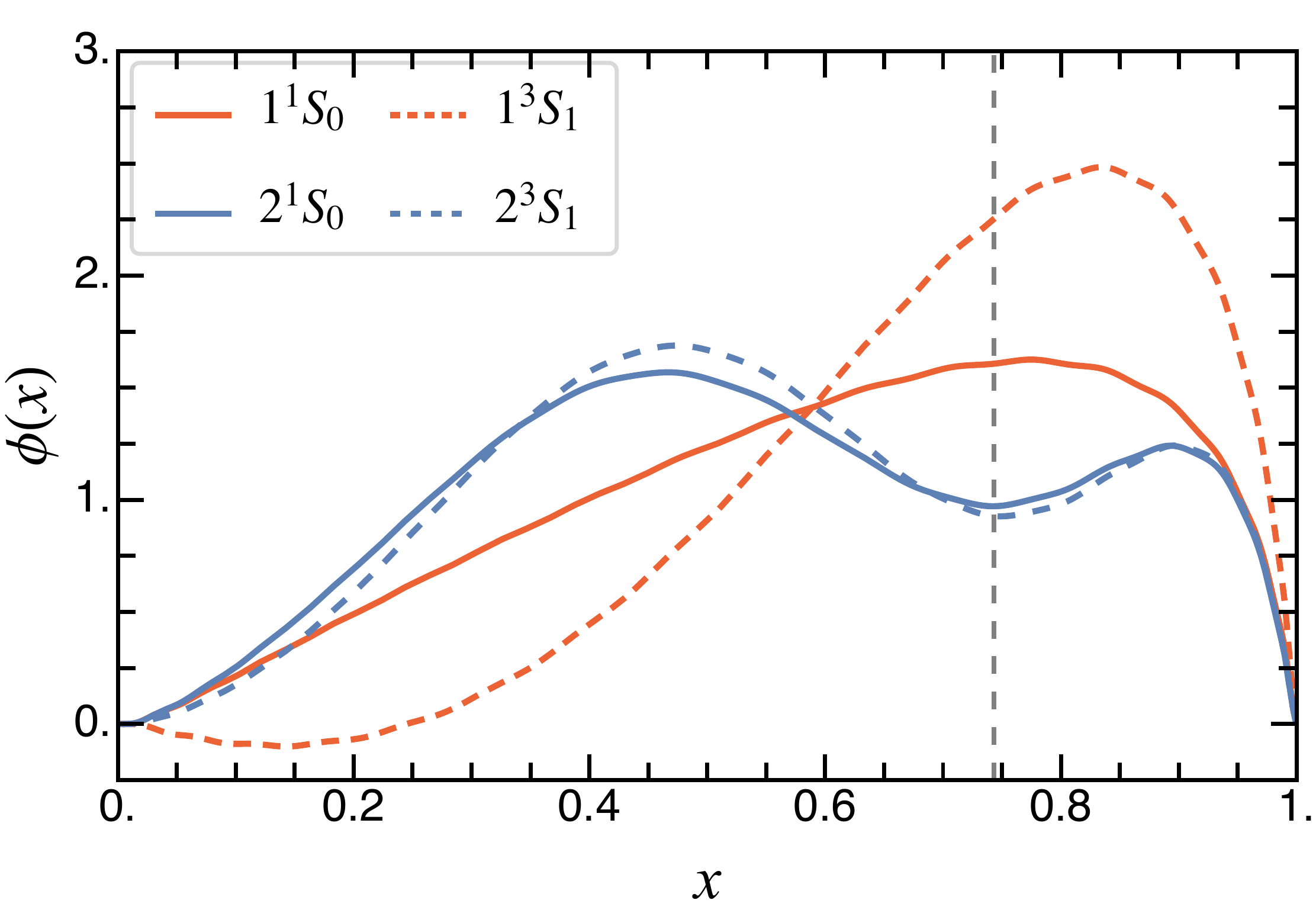} 
		\caption{$D$}
	\end{subfigure}
	\hspace{.5cm}
	\begin{subfigure}{0.45\textwidth}
		\centering
		\includegraphics[scale=0.35]{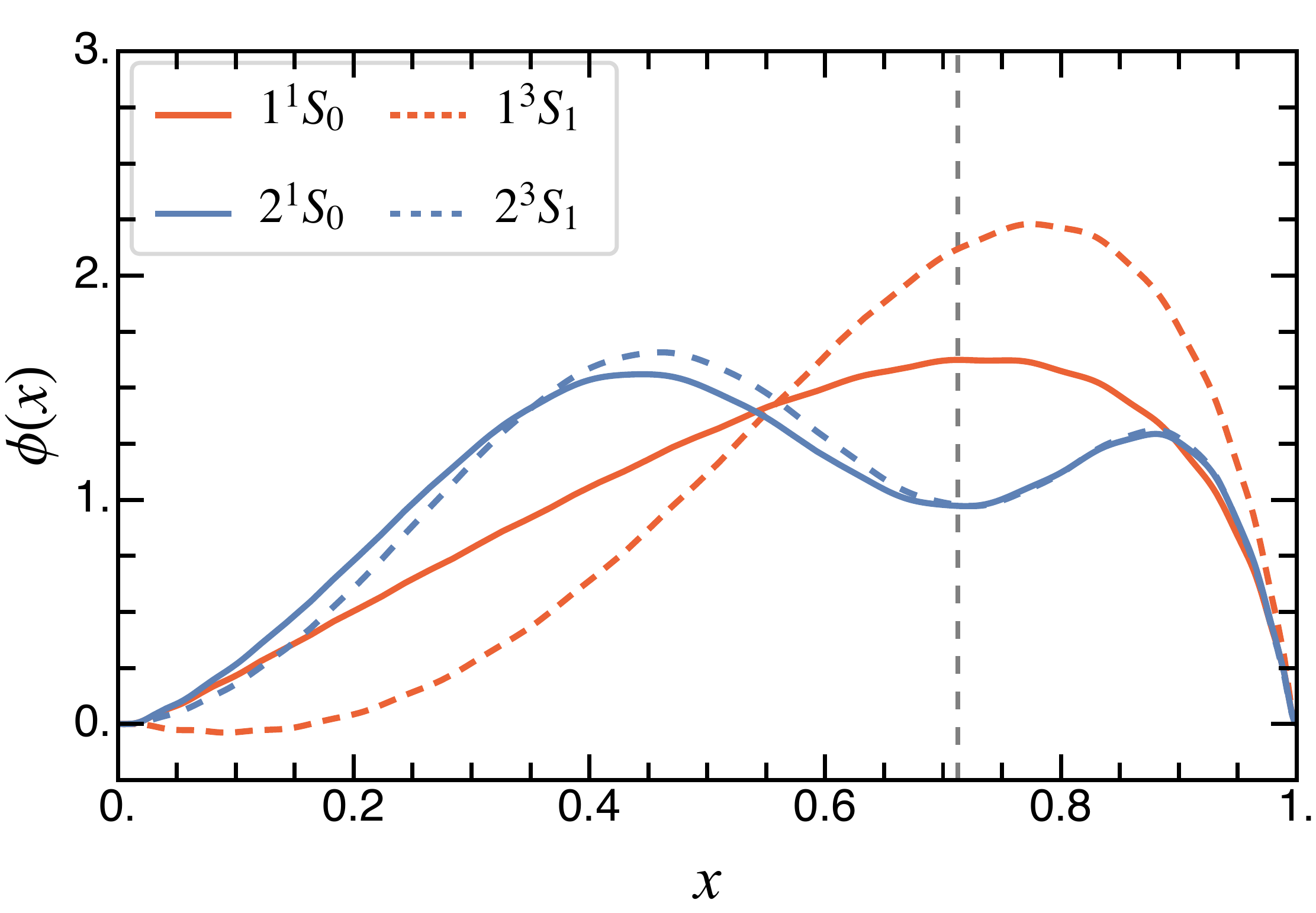}
		\caption{$D_s$}
	\end{subfigure} \\
	\begin{subfigure}{0.45\textwidth}
		\centering
		\includegraphics[scale=0.35]{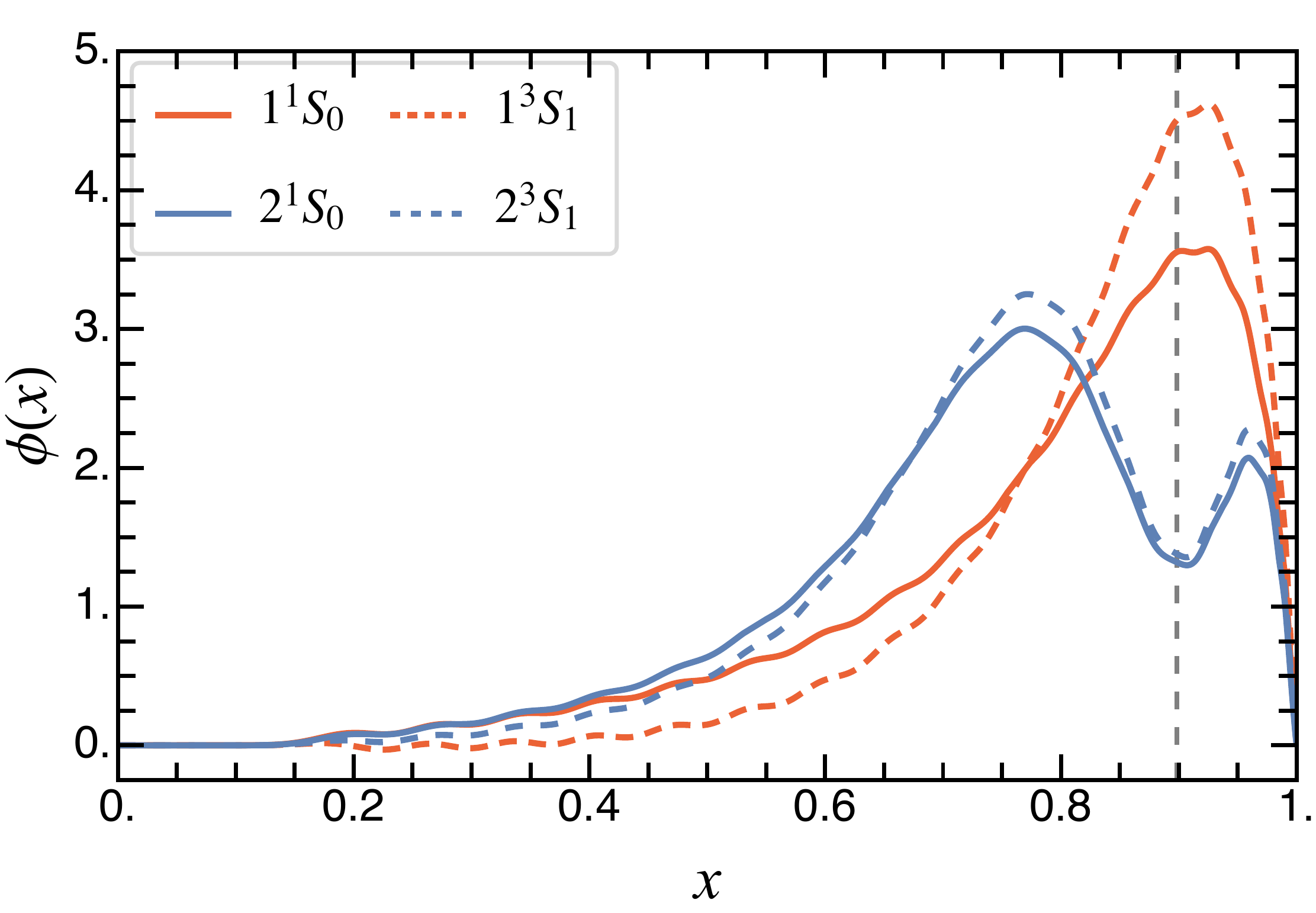} 
		\caption{$B$}
	\end{subfigure}
	\hspace{.5cm}
	\begin{subfigure}{0.45\textwidth}
		\centering
		\includegraphics[scale=0.35]{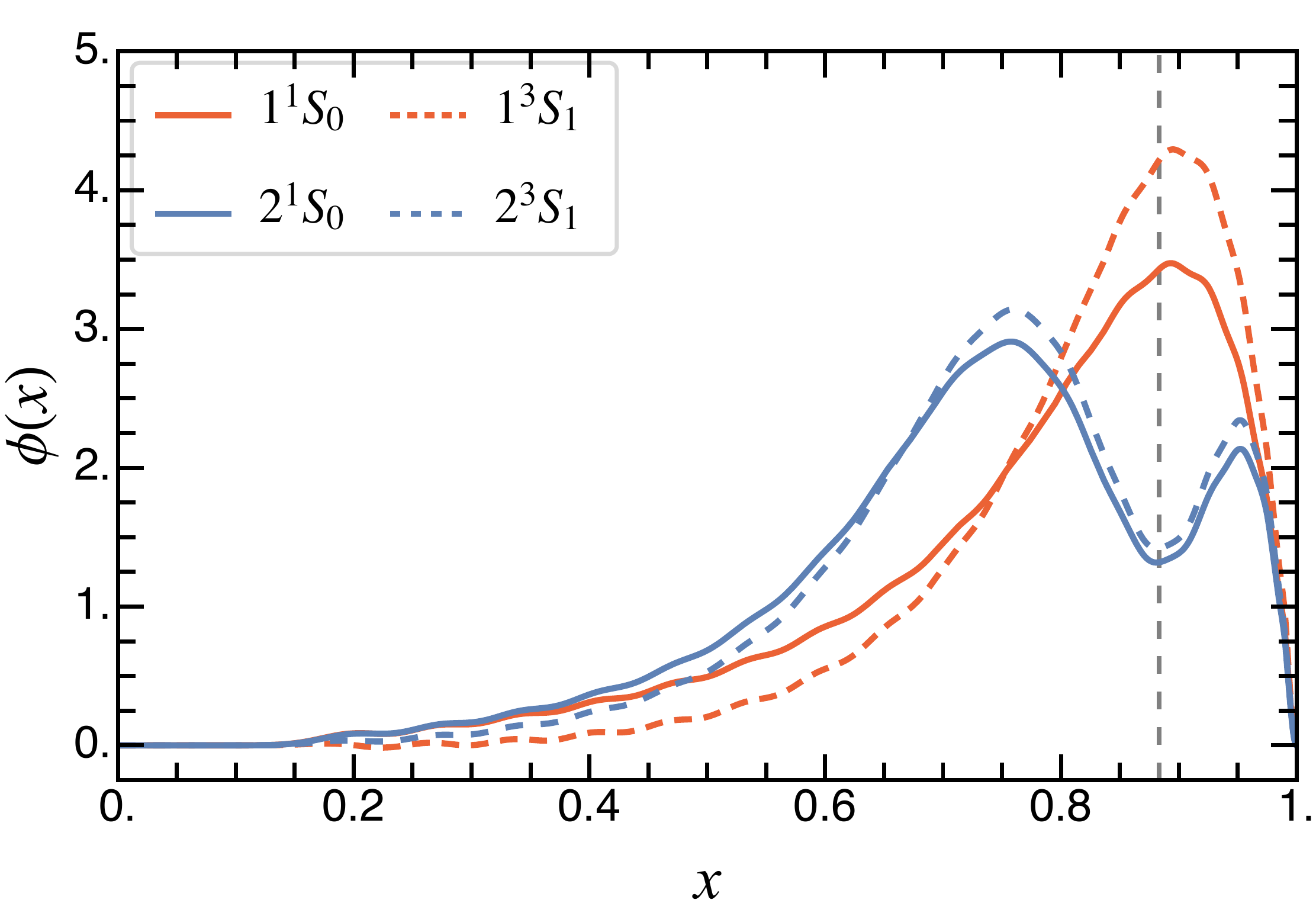}
		\caption{$B_s$}
	\end{subfigure} \\	
	\caption{\label{fig7} DAs of pseudoscalar (solid curves) and vector (dashed curves) states of the heavy-light systems at $N_\text{max} = L_\text{max} = 32$. The vertical dashed gray lines is at $x=m_q/(m_q+m_{\bar{q}})$.}
	\vspace*{1cm}
\end{figure*}

Another light-cone distribution we investigate is the DA. DA controls the exclusive process at large momentum transfer that is defined from the lightlike vacuum-to-meson matrix elements. In the LFWF representation, it can be written as~\cite{PhysRevD.22.2157}
\begin{equation}
\begin{aligned}
\frac{f_{P,V}}{2\sqrt{2N_c}} \phi_{P,V}(x) =\frac{1}{\sqrt{x(1-x)}}\int^{\lesssim \mu^2} \frac{\dd^2 {k}_\perp}{2(2\pi)^3} \psi^{(m_J=0)}_{\uparrow\downarrow\mp\downarrow\uparrow}(x, \vec{k}_\perp),
\end{aligned}
\end{equation}
where $f_{P(V)}$ are the decay constants~\footnote{ Details of decay constant will be discussed in Sec.~\ref{subs5}.} of the pseudoscalar (vector) state, and is associated with the minus (plus) sign in the subscript of the wave function. Like in the PDF, we have the UV cutoff $\mu$ taken as $\mu \approx \kappa\sqrt{N_\text{max}}$ in the basis representation. In these definitions, DAs are normalized to unity when one uses the same basis cutoff for the LFWFs and decay constants.
We compare the DAs of four heavy-light systems in Fig.~\ref{fig7}. 

Noticeably, DAs share some features with the PDFs: the DA spreads wider along $x$ in the lighter system as it is more relativistic; dips in $2^1S_0$ and $2^3S_1$ states reflect their character as radial excitations; the discrepancy between pseudoscalar and vector which is caused by the different spin configuration mixing appears more significant in lower exited states; and lastly, the asymmetries shown in the heavy-light meson DAs. However, different from PDFs, the DAs of the ground state ($1^1S_0$) peaked approximately at the quark mass fraction $x=m_q/(m_q+m_{\bar{q}})$, which is in accordance with the other unequal mass system $B_c$~\cite{PhysRevD.98.114038}. 
Noticeable wiggles that appear in DAs especially of $B$ and $B_s$ are due to the limited range of basis spaces employed. Similar patterns are also found in the light system~\cite{PhysRevC.99.035206}, and are understood to be resolved by increasing the basis size.
We find interesting similarities as well as differences of our DAs with those from the Dyson-Schwinger equation approach~\cite{BINOSI2019257}. For example, the DAs for our $D$ and $D_s$ ground states are more widely spread in $x$. On the other hand, the peak heights of our DAs for $B$ and $B_s$ are very similar to the corresponding DAs in Ref.~\cite{BINOSI2019257}.

\subsection{Decay Constants \label{subs5}}	
The decay constants provide important information of the internal structure of the mesons.
% and how that internal structure changes with different quark constituents.
$f_P$ and $f_V$ are the decay constants of the pseudoscalar and vector mesons, which characterize the strength of the matrix elements of the electroweak current between the corresponding meson and the vacuum. By definition, they are,
\begin{equation}
\begin{aligned}
\mel{0}{\overline{\psi}\gamma^\mu\gamma_5 \psi}{P(p)} & = \imag p^\mu f_P, \\
\mel{0}{\overline{\psi}\gamma^\mu \psi}{V(p, m_J)} & = \epsilon^\mu_{m_J} M_V f_V,
\end{aligned}
\end{equation} 
where $p^\mu$ is the four-momentum of the meson, $\epsilon^\mu_{m_J}$ and $M_V$ are the polarization vector and mass of the vector meson, respectively.
These decay constants correspond to the LFWFs at the origin in coordinate space, and can be calculated by integrating the LFWFs as follows~\cite{PhysRevD.96.016022,PhysRevD.98.114038}, 
\begin{equation}
f_{P,V} = 2\sqrt{2 N_c} \int_0^1 \frac{\dd x }{2\sqrt{x(1-x)}} \int \frac{\dd ^2 k_\perp}{(2\pi)^3}\psi^{(m_J=0)}_{\uparrow\downarrow\mp\downarrow\uparrow}(x, \vec{k}_\perp).
\end{equation}
We calculate the decay constants of the ground state pseudoscalar and vector of the heavy-light mesons, and compare with the experiments~\cite{PhysRevD.78.052003,Zupanc2013} and other published approaches~\cite{PhysRevD.98.074512,  PhysRevD.86.034506,aoki2019flag,PhysRevD.96.034524,PhysRevLett.110.222003,PhysRevD.91.114509,PhysRevD.60.074501,Becirevic2012,Dimopoulos2012,PhysRevD.91.054502,PhysRevD.85.114506,PhysRevD.91.054507,2014231,PhysRevD.86.054510,PhysRevD.88.014015,LUCHA201182,LUCHA201412,NARISON20131321,Wang2015,Yang2012,PhysRevD.99.093002,EBERT200693,CVETIC200484,PhysRevD.81.054022}. 
As mentioned earlier, the basis cutoff $N_\text{max}$ is associated with the UV regulator by $\Lambda_\text{UV} \approx \kappa \sqrt{N_\text{max}}$. 
For calculating the decay constants, we employ different $N_\text{max}$ values for different systems so that $\Lambda_\text{UV} \approx 0.85 (m_q + m_{\bar{q}})$ in accordance with our previous work of heavy quarkonia~\cite{PhysRevD.96.016022}: we apply $N_\text{max} = 8$ for $D$ and $D_s$, while $N_\text{max} =16$ for $B$ and $B_s$. Uncertainties quoted for BLFQ are used to show the basis cutoff sensitivity, where we take $\Delta f =\abs{ f(N_\text{max}=8) -  f(N_\text{max}=16)}$
for all heavy-light systems. The choice of $N_\text{max}$ is a result of compromise between the needs for better basis resolution and a lower UV scale owing to the omitted radiative corrections.

The difference of the decay constants of pseudoscalar and vector heavy-light mesons is due to differences in the internal spin configurations. 
%	When we compare the ratio of $f_P/f_V$, it turns out that there is significant tension among different methods. 
There is a significant tension among results obtained from different methods. Within the same approach, Lattice QCD for example, the tension still exists among the results reported by different groups~\cite{PhysRevLett.110.222003,PhysRevD.91.114509,PhysRevD.60.074501}. 
In this work, $f_P/f_V$ is greater than 1 for all heavy-light mesons, whereas the QCD sum rules (QCDSR) and relativistic quark model (RQM) quoted here favor a value less than 1. Details have been discussed in Refs.~\cite{PhysRevD.91.114509,Becirevic:2014kaa,Lucha:2014hqa,NARISON2016143}.
In addition, we check the ratio for the light mesons $f_{K}/f_{K^*}$ based on the same model, and obtain a result slightly smaller than 1, which agrees with experiments and also suggests this ratio is not systematically larger than 1 as seen in heavy systems. 
However, more work is needed to bring in additional physics, including the role of chiral symmetry~\cite{MARIS2006136,PhysRevC.56.3369}, before one can extend this approach to light systems~\cite{WenyangQian,PhysRevC.99.035206}.
%	While for various Lattice collaborations, the ratio differs within systems as well as across approaches. 
%	In BLFQ, the decay constant, especially the ratio of $f_P/f_V$, may vary when introduce the chiral dynamics considered the light quark effect, which remains to be investigated.

\begin{table*}
	\begin{tabular}{ccccccccccc}
		\hline\noalign{\smallskip}
		(MeV)&  $f_{D}$  &   $f_{D^*}$   & $f_{D_s}$  &  $f_{D^*_s}$  & $f_{B}$ &  $f_{B^*}$  &   $f_{B_s}$  &  $f_{B^*_s}$ & Ref.  \\
		\noalign{\smallskip}\hline\noalign{\smallskip}
		BLFQ 	 &$295(63)$& $281(20)$ & $313(67)$ & $306(39)$& $233(50)$&$202(32)$&$259(54)$& $230(36)$ &	this work\\
		Exp. &$205.8(8.9)$& ...& $255.5(6.6)$ & ... &...&...&...&... &\cite{PhysRevD.78.052003,Zupanc2013} \\
		%			Lattice-1 &$212.7(0.6)$& ...& $249.9(0.4)$ & ... &$189.4(1.4)$&...&$230.7(1.3)$&... &\cite{PhysRevD.98.074512} \\
		%			%			 &... & ... & ... & ...& $191(9)$&...&$228(10)$&... &  \cite{PhysRevD.86.034506}\\
		%			-2&... & ... & ... & ...& $186(4)$&$175(6)$&$224(4)$&$213(7)$&  \cite{PhysRevLett.110.222003,PhysRevD.91.114509}\\
		%			-3&$212(8)$& $278(16)$ &$248(6)$&  $311(9)$ &$195(12)$&...&$232(10)$&... &\cite{Becirevic2012,Dimopoulos2012}	\\
		FLAG $N_f=4$& $212.0(0.7)$ & ... &$249.9(0.5)$& ...&$ 190.0(1.3)$&...&$230.3(1.3)$	&...&\cite{aoki2019flag}		\\
		FLAG $N_f=3$& $ 209.0(2.4)$ & ...& $ 248.0(1.6)$ & ...& $ 192.0(4.3)$& ...&$228.4(3.7)$&... &\cite{aoki2019flag}	\\
		Lattice &...& $223.5(8.4)$& ...& $268.8(6.6)$ &...&$185.9(7.2)$&...&$223.1(5.4)$ &\cite{PhysRevD.96.034524} \\
		%			 &... & ... & ... & ...& $195.6(16.2)$&...&$235.4(12.3)$&... &  \cite{PhysRevD.91.054502}\\
		%			 &$207.4(3.8)$& ...& $247.2(4.1)$ & ... &...&...&...&... &\cite{PhysRevD.91.054507} \\
		%			 &$202.3(3.4)$& ...& $258.7(3.1)$ & ... &...&...&...&... &\cite{2014231}\\
		%			 &$208.3(3.5)$& ...& $246(3.6)$ & ... &...&...&...&... &\cite{PhysRevD.86.054510}\\
		%			 &$211(18)$& $245(20)$& $231(14)$ &$272(26)$ &$179(38)$&$196(46)$&$204(39)$&$229(46)$ &\cite{PhysRevD.60.074501}\\
		QCDSR-1 &$201_{-13}^{+12}$& $242_{-12}^{+20}$& $238_{-23}^{+13}$ & $314_{-14}^{+19}$ &$207_{-9}^{+17}$&$210_{-12}^{+10}$&$242_{-12}^{+17}$&$267_{-20}^{+14}$ &\cite{PhysRevD.88.014015} \\
		-2&$206.3(8.9)$& $252.2(22.7)$& $245(16.3)$ & $305.5(27.3)$ &...&...&...&... &\cite{LUCHA201182,LUCHA201412}\\
		%			-3&$204(6)$&...& $246(6)$ & ... &$206(7)$&...&$234(5)$&... &\cite{NARISON20131321} \\
		-3&$208(10)$&$263(21)$& $240(10)$ & $308(21)$ &$194(15)$&$213(18)$&$231(16)$&$255(19)$ &\cite{Wang2015} \\
		%			QM-1 &$208(21)$&...&$256(26)$& ... &$198(14)$&...&$237(17)$&...& \cite{Yang2012}\\			 
		RQM&...&...&...& ... &$210(10)$&$223(16)$&$229(11)$&$242(17)$ &\cite{PhysRevD.99.093002}\\
		LFQM&...&...&$264.5(17.5)$& ... &...&...&$270.0(42.8)$&...&\cite{PhysRevD.81.054022}\\
		%						 &$234$&...& $268$ & ... &$189$&...&$218$&... &\cite{EBERT200693} \\
		%						 &$230(25)$&...& $248(27)$ & ... &$196(29)$&...&$216(32)$&... &\cite{CVETIC200484} \\
		\hline\noalign{\smallskip}
	\end{tabular}
	\caption{The decay constant of ground state pseudoscalar and vector states of four heavy-light systems. We list our results (BLFQ) and compare to existing experimental results, and other theoretical methods: Lattice QCD and the averaged results summarized in Flavour Lattice Averaging Group Review 2019 (FLAG), QCD sum rules (QCDSR), relativistic quark model (RQM), and light-front quark model (LFQM).\label{tb2}}
	\vspace*{1cm} 
\end{table*}

\begin{figure*}
	\centering
	\hspace{-1.5cm}
	\begin{subfigure}{0.45\textwidth}
		\centering
		\includegraphics[scale=0.4]{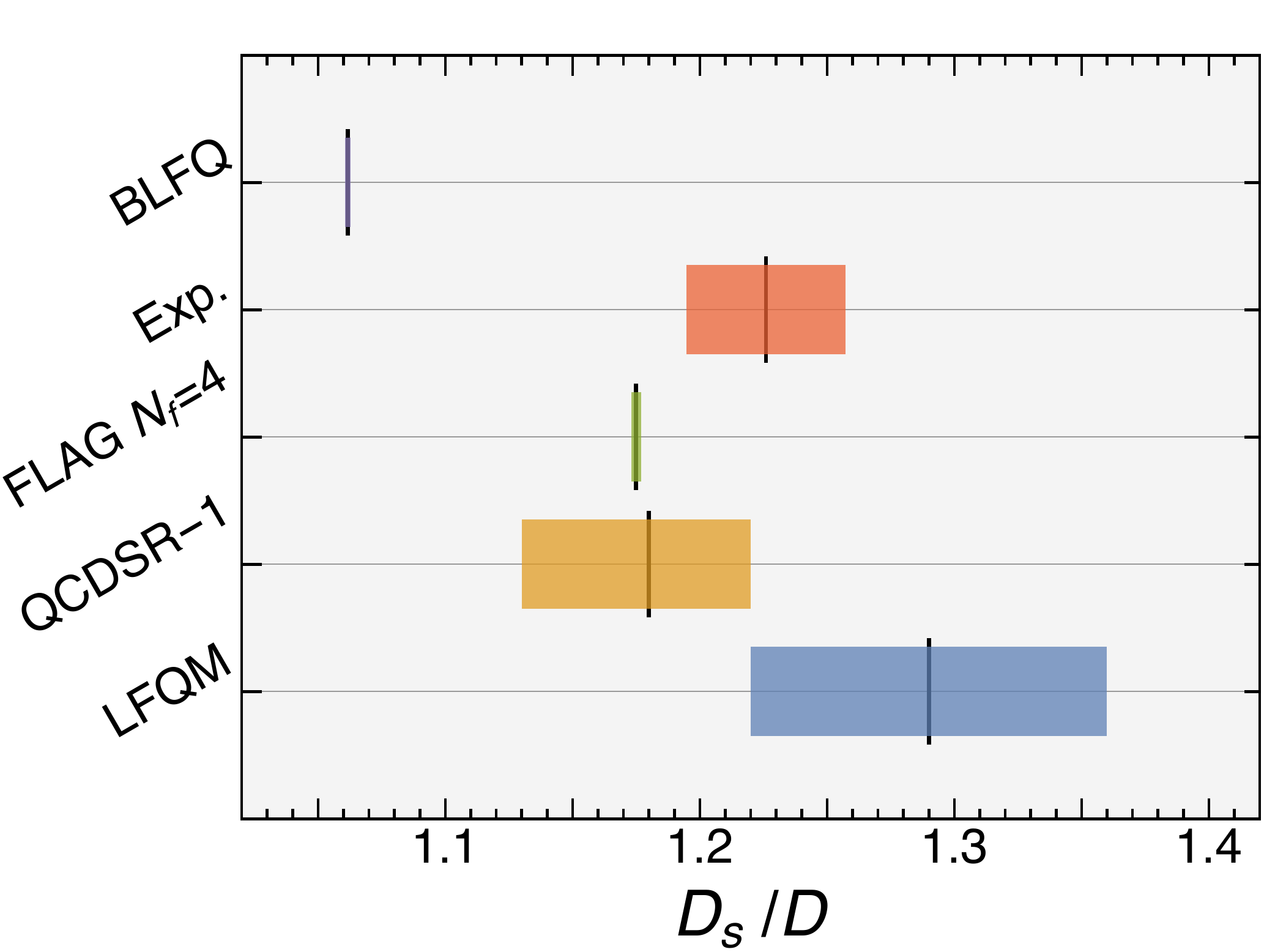} 
		%	\caption{$D^0$}
	\end{subfigure}
	\hspace{1cm}
	\begin{subfigure}{0.45\textwidth}
		\centering
		\includegraphics[scale=0.4]{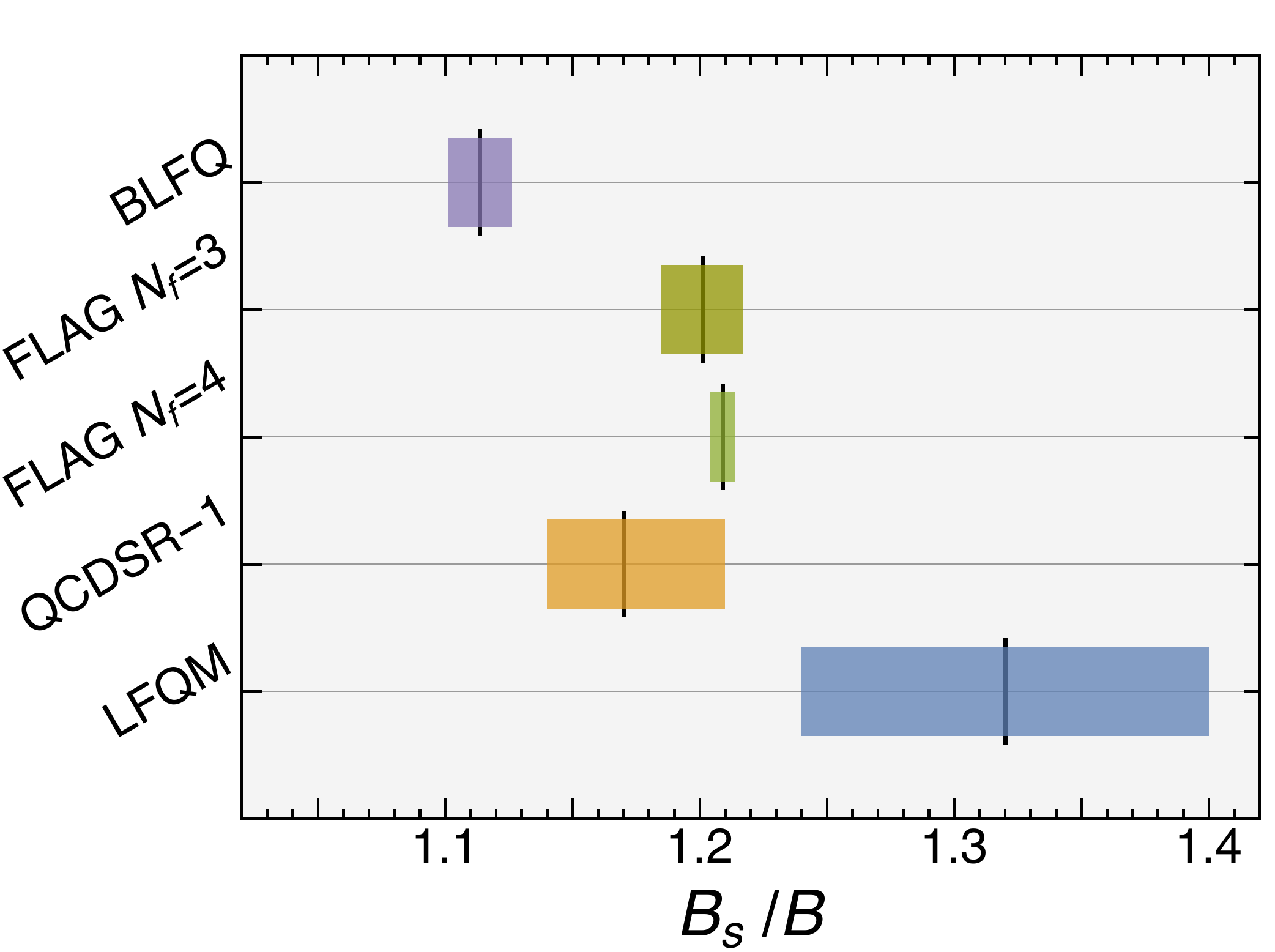}
		%	\caption{$D_s$}
	\end{subfigure}
	\caption{\label{fig3}Comparison of the ratios $\mathcal{R}(f_{D_s}/f_{D})$ and $\mathcal{R}(f_{B_s}/f_{B})$ with other approaches~\cite{aoki2019flag,PhysRevD.88.014015,PhysRevD.81.054022} and experimental average~\cite{PhysRevD.98.030001}. See Table~\ref{tb2} for the specific values and their sources for each result plotted here. The vertical black lines indicate the central value of the ratios, while rectangles are the corresponding uncertainties.}
	\vspace*{1cm}
\end{figure*}

The ratios of the decay constants for the mesons which consist of only one different flavored quark is another interesting quantity. These quantities can reduce some of the systematic bias in theoretical and experimental approaches.
The ratios $\mathcal{R}(f_{B_s}/f_{B})$ and $\mathcal{R}(f_{D_s}/f_{D})$ are presented in Fig.~\ref{fig3}, with a comparison between results from other methods and from experiments. The vertical black lines indicate the central values of the ratio, and rectangles indicate the corresponding uncertainties. 
%	Note that the ratio we present for the experimental average is from a combination of sources \cite{PhysRevD.78.052003,Zupanc2013}, and the corresponding uncertainties are treated independently.
For the BLFQ results, the central lines are the ratios calculated with $N_\text{max} = 8$ for $D$ and $D_s$, $N_\text{max} =16$ for $B$ and $B_s$, respectively. Meanwhile the uncertainty is quoted as $\Delta_\mathcal{R} = \abs{\mathcal{R}(N_\text{max}=8)-\mathcal{R}(N_\text{max}=16)}$, because $\Delta f$ is strongly correlated. Unlike the decay constant itself, the ratio does not have a monotonic dependence on the basis size. Among the several methods, ratios from BLFQ are smaller than those from other approaches by $5-18\%$.  We mention that the decay constant of a pseudoscalar  meson, taking the flavor symmetry breaking into account, is discussed in Ref.~\cite{osti_7335315}.

\section{Summary and Outlook}			
\label{sec3}
In this work, a model consisting of a holographic confining interaction, a longitudinal confinement and a one-gluon exchange is applied to mesons containing a light quark and is solved within the BLFQ method. The mass spectroscopy shows reasonable agreement with the experimental measurements and Lattice QCD calculations. This extends the previously successful application of the same model to heavy mesons. In this work, we fitted two parameters $m_\mathsf{u/d}$ and $m_\mathsf{s}$, and all those parameters such as $m_\mathsf{c}$ and $\kappa_\mathsf{cc}$ are universal among these systems. 
%we tested the model for the heavy-light mesons within the BLFQ approach. This model has first been adopted for heavy mesons and achieved plausible agreement with experiments, as well as with other methods. The same effective Hamiltonian for heavy mesons without adding additional interaction terms is adopted for the meson containing a light quark, and it also shows reasonable agreement at the mass spectrum. 
%	More importantly, as a advantage of BLFQ, it provides fruitful prediction on excited meson states. 
%By far, another block of BLFQ on mesons has been validated. 
Without incorporating a chiral effective term in the Hamiltonian, we still obtain reasonable hyperfine splitting between the ground state pseudoscalar and vector, which suggests that the chiral dynamics is not dominant in the spectra of heavy-light mesons as much as in the pure light systems.

We investigate the LFWFs of the heavy-light systems and found non-trivial distribution of peaks along $x$. 
Due to this attribute of the LFWFs, the light-cone distributions such as PDF and DA, calculated by integrating the wave function, will be effected. Therefore we studied the PDFs and DAs of the heavy-light systems, and noticed that the peak of these functions, were not always locate at the quark mass fraction as one might naively expect. In particular, the peak of ground state DA is located close to the $x=m_q/(m_q+m_{\bar{q}})$, but that of PDF is located well above $x=m_q/(m_q+m_{\bar{q}})$.

We calculated the decay constants and the ratios which are also of great experimental and theoretical interest.
The values of decay constant vary with the basis cutoffs due to the lack of radiative corrections in our model. Nevertheless, the ratio of heavier to lighter meson decay constants remains relatively stable.
Even though the absence of chiral dynamics seems not to spoil the mass spectrum of heavy-light mesons, one could add an effective chiral potential in the Hamiltonian~\cite{PhysRevC.99.035206} in a future work, in order to address the decay constant ratio of vector to pseudoscalar more rigorously and investigate the role of chiral effects. In the mean time, incorporating the self-energy correction by including higher Fock sectors is another direction of improvement, which could reduce the sensitive dependence on basis size of the decay constant.

%With the DGLAP tool which is tested in BLFQ, we can improve its resolution to experimental required level.

The fitted masses of light quarks of this work agree well with the fits from light mesons~\cite{WenyangQian}, both of which provide $m_\mathsf{u}/m_\mathsf{d} \sim 550$ MeV in the BLFQ approach. The obtained LFWFs for a wide range of meson systems provide opportunities to explore a broader class of observables.
Notable examples include the dilepton decays, diphoton decays, radiative transitions, and weak decays. 
%, such as the radiative transitions between different states, weak decays between mesons.

\section{Acknowledgments}			
We wish to thank S. Jia, M. Li, W. Qian, and A. Yu for fruitful discussions. S.T. and J.P.V. wish to thank Dr. Soeren Prell for valuable advice.
This work was supported in part by the U.S. Department of Energy under Grants No. DE-FG02-87ER40371 and No. DE-SC0018223 (SciDAC-4/NUCLEI). Computational resources were provided by the National Energy Research Supercomputer Center, which is supported by the Office of Science of the U.S. Department of Energy under Contract No. DE-AC02-05CH11231.

\bibliographystyle{unsrt}
\bibliography{heavy_light_ref}

\begin{thebibliography}{10}

\bibitem{PhysRevC.81.035205}
J.~P. Vary, H.~Honkanen, Jun Li, P.~Maris, S.~J. Brodsky, A.~Harindranath,
  G.~F. de~T\'eramond, P.~Sternberg, E.~G. Ng, and C.~Yang.
\newblock Hamiltonian light-front field theory in a basis function approach.
\newblock {\em Phys. Rev. C}, 81:035205, Mar 2010.

\bibitem{PhysRevD.96.016022}
Yang Li, Pieter Maris, and James~P. Vary.
\newblock Quarkonium as a relativistic bound state on the light front.
\newblock {\em Phys. Rev. D}, 96:016022, Jul 2017.

\bibitem{PhysRevD.98.114038}
Shuo Tang, Yang Li, Pieter Maris, and James~P. Vary.
\newblock ${B}_{c}$ mesons and their properties on the light front.
\newblock {\em Phys. Rev. D}, 98:114038, Dec 2018.

\bibitem{ISGUR1989113}
Nathan Isgur and Mark~B. Wise.
\newblock Weak decays of heavy mesons in the static quark approximation.
\newblock {\em Physics Letters B}, 232(1):113 -- 117, 1989.

\bibitem{PhysRevD.48.4370}
M.~A. Nowak, M.~Rho, and I.~Zahed.
\newblock Chiral effective action with heavy-quark symmetry.
\newblock {\em Phys. Rev. D}, 48:4370--4374, Nov 1993.

\bibitem{PhysRevD.49.409}
William~A. Bardeen and Christopher~T. Hill.
\newblock Chiral dynamics and heavy quark symmetry in a solvable toy
  field-theoretic model.
\newblock {\em Phys. Rev. D}, 49:409--425, Jan 1994.

\bibitem{NIELSEN201041}
Marina Nielsen, Fernando~S. Navarra, and Su~Houng Lee.
\newblock New charmonium states in qcd sum rules: A concise review.
\newblock {\em Physics Reports}, 497(2):41 -- 83, 2010.

\bibitem{MARIS2006136}
P.~Maris and P.C. Tandy.
\newblock {QCD} modeling of hadron physics.
\newblock {\em Nuclear Physics B - Proceedings Supplements}, 161:136 -- 152,
  2006.
\newblock Proceedings of the Cairns Topical Workshop on Light-Cone QCD and
  Nonperturbative Hadron Physics.

\bibitem{PhysRevLett.113.162001}
R.~Aaij, B.~Adeva, M.~Adinolfi, A.~Affolder, Z.~Ajaltouni, S.~Akar,
  J.~Albrecht, F.~Alessio, M.~Alexander, S.~Ali, and et. al.
\newblock Observation of overlapping {S}pin-1 and {S}pin-3
  $\overline{D}^{0}{K}^{\ensuremath{-}}$ resonances at mass $2.86\text{ }\text{
  }\mathrm{GeV}/{c}^{2}$.
\newblock {\em Phys. Rev. Lett.}, 113:162001, Oct 2014.

\bibitem{Aaij2015}
{The LHCb collaboration}, R.~Aaij, B.~Adeva, M.~Adinolfi, A.~Affolder,
  Z.~Ajaltouni, S.~Akar, J.~Albrecht, F.~Alessio, M.~Alexander, S.~Ali, and et.
  al.
\newblock Precise measurements of the properties of the ${B}_1(5721)^{0,+}$ and
  ${B}_2^*(5747)^{0,+}$ states and observation of ${B}^{+,0}$, $\pi^{-,+}$ mass
  structures.
\newblock {\em Journal of High Energy Physics}, 2015(4):24, Apr 2015.

\bibitem{PhysRevLett.10.531}
Nicola Cabibbo.
\newblock Unitary symmetry and leptonic decays.
\newblock {\em Phys. Rev. Lett.}, 10:531--533, Jun 1963.

\bibitem{10.1143/PTP.49.652}
Makoto Kobayashi and Toshihide Maskawa.
\newblock {CP-Violation in the Renormalizable Theory of Weak Interaction}.
\newblock {\em Progress of Theoretical Physics}, 49(2):652--657, 02 1973.

\bibitem{NARISON2008308}
Stephan Narison.
\newblock $|{V}_{cd}|$, $|{V}_{cs}|$ and $f_{D_{(s)}}$ from (semi-) leptonic
  ${D}_{(s)}$-decays: Signals of new physics?
\newblock {\em Physics Letters B}, 668(4):308 -- 311, 2008.

\bibitem{PhysRevLett.102.081601}
Guy~F. de~T\'eramond and Stanley~J. Brodsky.
\newblock Light-front holography: A first approximation to {QCD}.
\newblock {\em Phys. Rev. Lett.}, 102:081601, Feb 2009.

\bibitem{LI2016118}
Yang Li, Pieter Maris, Xingbo Zhao, and James~P. Vary.
\newblock Heavy quarkonium in a holographic basis.
\newblock {\em Physics Letters B}, 758:118 -- 124, 2016.

\bibitem{BRODSKY20151}
Stanley~J. Brodsky, Guy~F. de~T\'eramond, Hans~G{\"u}nter Dosch, and Joshua
  Erlich.
\newblock Light-front holographic {QCD} and emerging confinement.
\newblock {\em Physics Reports}, 584:1 -- 105, 2015.
\newblock Light-front holographic QCD and emerging confinement.

\bibitem{WenyangQian}
Wenyang Qian, Shaoyang Jia, Yang Li, and James~P. Vary.
\newblock {Light mesons within the basis light-front quantization framework}.
\newblock In preparation.

\bibitem{PhysRevC.99.035206}
Shaoyang Jia and James~P. Vary.
\newblock Basis light front quantization for the charged light mesons with
  color singlet {N}ambu-{J}ona-{L}asinio interactions.
\newblock {\em Phys. Rev. C}, 99:035206, Mar 2019.

\bibitem{PhysRevC.56.3369}
Pieter Maris and Craig~D. Roberts.
\newblock $\ensuremath{\pi}$- and ${K}$-meson {B}ethe-{S}alpeter amplitudes.
\newblock {\em Phys. Rev. C}, 56:3369--3383, Dec 1997.

\bibitem{PhysRevD.98.030001}
M.~Tanabashi, K.~Hagiwara, K.~Hikasa, K.~Nakamura, Y.~Sumino, F.~Takahashi,
  J.~Tanaka, K.~Agashe, G.~Aielli, C.~Amsler, and et~al.
\newblock Review of particle physics.
\newblock {\em Phys. Rev. D}, 98:030001, Aug 2018.

\bibitem{PhysRevD.83.014506}
Eric~B. Gregory, Christine T.~H. Davies, Iain~D. Kendall, Jonna Koponen, Kit
  Wong, Eduardo Follana, Elvira G\'amiz, G.~Peter Lepage, Eike~H. M\"uller,
  Heechang Na, and Junko Shigemitsu.
\newblock Precise ${B}$, ${B}_{s}$, and ${B}_{c}$ meson spectroscopy from full
  lattice {QCD}.
\newblock {\em Phys. Rev. D}, 83:014506, Jan 2011.

\bibitem{PhysRevD.86.094510}
R.~J. Dowdall, C.~T.~H. Davies, T.~C. Hammant, and R.~R. Horgan.
\newblock Precise heavy-light meson masses and hyperfine splittings from
  lattice {QCD} including charm quarks in the sea.
\newblock {\em Phys. Rev. D}, 86:094510, Nov 2012.

\bibitem{LANG201517}
C.B. Lang, Daniel Mohler, Sasa Prelovsek, and R.M. Woloshyn.
\newblock Predicting positive parity ${B}_s$ mesons from lattice {QCD}.
\newblock {\em Physics Letters B}, 750:17 -- 21, 2015.

\bibitem{PhysRevD.92.094508}
Martin Kalinowski and Marc Wagner.
\newblock Masses of ${D}$ mesons, ${D}_{s}$ mesons, and charmonium states from
  twisted-mass lattice {QCD}.
\newblock {\em Phys. Rev. D}, 92:094508, Nov 2015.

\bibitem{kou2018belle}
E.~Kou, P.~Urquijo, W.~Altmannshofer, F.~Beaujean, G.~Bell, M.~Beneke, I.~I.
  Bigi, F.~Bishara~M. Blanke, C.~Bobeth, M.~Bona, N.~Brambilla, and et~al.
\newblock The {B}elle {II} physics book.
\newblock 2018.

\bibitem{Jia:2018hxd}
Shaoyang Jia and James~P. Vary.
\newblock {Parton distribution functions from scalar light front parton gas
  model}.
\newblock 2018.

\bibitem{Maris:2020wew}
Pieter Maris, Shaoyang Jia, Meijian Li, Yang Li, Shuo Tang, and James~P. Vary.
\newblock {On the light-front wave functions of quarkonia}.
\newblock In {\em {Light Cone 2019 (LC2019) Palaiseau, France, September 16-20,
  2019}}, 2020.

\bibitem{PhysRevLett.122.172001}
Jiangshan Lan, Chandan Mondal, Shaoyang Jia, Xingbo Zhao, and James~P. Vary.
\newblock Parton distribution functions from a light front hamiltonian and
  {QCD} evolution for light mesons.
\newblock {\em Phys. Rev. Lett.}, 122:172001, May 2019.

\bibitem{PhysRevD.22.2157}
G.~Peter Lepage and Stanley~J. Brodsky.
\newblock Exclusive processes in perturbative quantum chromodynamics.
\newblock {\em Phys. Rev. D}, 22:2157--2198, Nov 1980.

\bibitem{BINOSI2019257}
Daniele Binosi, Lei Chang, Minghui Ding, Fei Gao, Joannis Papavassiliou, and
  Craig~D. Roberts.
\newblock Distribution amplitudes of heavy-light mesons.
\newblock {\em Physics Letters B}, 790:257 -- 262, 2019.

\bibitem{PhysRevD.78.052003}
B.~I. Eisenstein, I.~Karliner, S.~Mehrabyan, N.~Lowrey, M.~Selen, E.~J. White,
  J.~Wiss, R.~E. Mitchell, M.~R. Shepherd, D.~Besson, and et~al.
\newblock Precision measurement of
  $\mathcal{B}({D}^{+}\ensuremath{\rightarrow}{\ensuremath{\mu}}^{+}\ensuremath{\nu})$
  and the pseudoscalar decay constant ${f}_{{D}^{+}}$.
\newblock {\em Phys. Rev. D}, 78:052003, Sep 2008.

\bibitem{Zupanc2013}
A.~Zupanc, I.~Adachi, H.~Aihara, K.~Arinstein, D.~M. Asner, T.~Aushev, A.~M.
  Bakich, A.~Bala, B.~Bhuyan, G.~Bonvicini, and et~al.
\newblock Measurements of branching fractions of leptonic and hadronic
  ${D}_s^+$ meson decays and extraction of the ${D}_s^+$ meson decay constant.
\newblock {\em Journal of High Energy Physics}, 2013(9):139, Sep 2013.

\bibitem{PhysRevD.98.074512}
A.~Bazavov, C.~Bernard, N.~Brown, C.~DeTar, A.~X. El-Khadra, E.~G\'amiz, Steven
  Gottlieb, U.~M. Heller, J.~Komijani, A.~S. Kronfeld, J.~Laiho, P.~B.
  Mackenzie, E.~T. Neil, J.~N. Simone, R.~L. Sugar, D.~Toussaint, and R.~S.
  Van~de Water.
\newblock ${B}$- and ${D}$-meson leptonic decay constants from four-flavor
  lattice {QCD}.
\newblock {\em Phys. Rev. D}, 98:074512, Oct 2018.

\bibitem{PhysRevD.86.034506}
Heechang Na, Chris~J. Monahan, Christine T.~H. Davies, Ron Horgan, G.~Peter
  Lepage, and Junko Shigemitsu.
\newblock ${B}$ and ${B}_{s}$ meson decay constants from lattice {QCD}.
\newblock {\em Phys. Rev. D}, 86:034506, Aug 2012.

\bibitem{aoki2019flag}
S.~Aoki, Y.~Aoki, D.~Becirevic, T.~Blum, G.~Colangelo, S.~Collins, M.~Della
  Morte, P.~Dimopoulos, S.~Dürr, H.~Fukaya, and et~al.
\newblock Flag review 2019.
\newblock 2019.

\bibitem{PhysRevD.96.034524}
V.~Lubicz, A.~Melis, and S.~Simula.
\newblock Masses and decay constants of ${D}_{(s)}^{*}$ and ${B}_{(s)}^{*}$
  mesons with ${N}_{f}=2+1+1$ twisted mass fermions.
\newblock {\em Phys. Rev. D}, 96:034524, Aug 2017.

\bibitem{PhysRevLett.110.222003}
R.~J. Dowdall, C.~T.~H. Davies, R.~R. Horgan, C.~J. Monahan, and J.~Shigemitsu.
\newblock ${B}$-meson decay constants from improved lattice nonrelativistic
  {QCD} with physical $u$, $d$, $s$, and $c$ quarks.
\newblock {\em Phys. Rev. Lett.}, 110:222003, May 2013.

\bibitem{PhysRevD.91.114509}
B.~Colquhoun, C.~T.~H. Davies, J.~Kettle, J.~Koponen, A.~T. Lytle, R.~J.
  Dowdall, and G.~P. Lepage.
\newblock ${B}$-meson decay constants: A more complete picture from full
  lattice {QCD}.
\newblock {\em Phys. Rev. D}, 91:114509, Jun 2015.

\bibitem{PhysRevD.60.074501}
D.~Becirevic, Ph. Boucaud, J.~P. Leroy, V.~Lubicz, G.~Martinelli, F.~Mescia,
  and F.~Rapuano.
\newblock Nonperturbatively improved heavy-light mesons: Masses and decay
  constants.
\newblock {\em Phys. Rev. D}, 60:074501, Aug 1999.

\bibitem{Becirevic2012}
Damir Be{\v{c}}irevi{\'{c}}, Vittorio Lubicz, Francesco Sanfilippo, Silvano
  Simula, and Cecilia Tarantino.
\newblock ${D}$-meson decay constants and a check of factorization in
  non-leptonic ${B}$-decays.
\newblock {\em Journal of High Energy Physics}, 2012(2):42, Feb 2012.

\bibitem{Dimopoulos2012}
P.~Dimopoulos, R.~Frezzotti, G.~Herdoiza, V.~Lubicz, C.~Michael, D.~Palao,
  G.~C. Rossi, F.~Sanfilippo, A.~Shindler, S.~Simula, C.~Tarantino, and
  M.~Wagner.
\newblock Lattice {QCD} determination of $m_b$, $f_{B}$ and $f_{B_s}$ with
  twisted mass wilson fermions.
\newblock {\em Journal of High Energy Physics}, 2012(1):46, Jan 2012.

\bibitem{PhysRevD.91.054502}
N.~H. Christ, J.~M. Flynn, T.~Izubuchi, T.~Kawanai, C.~Lehner, A.~Soni, R.~S.
  Van~de Water, and O.~Witzel.
\newblock ${B}$-meson decay constants from $2+1$-flavor lattice {QCD} with
  domain-wall light quarks and relativistic heavy quarks.
\newblock {\em Phys. Rev. D}, 91:054502, Mar 2015.

\bibitem{PhysRevD.85.114506}
A.~Bazavov, C.~Bernard, C.~M. Bouchard, C.~DeTar, M.~Di~Pierro, A.~X.
  El-Khadra, R.~T. Evans, E.~D. Freeland, E.~G\'amiz, Steven Gottlieb, and
  et~al.
\newblock ${B}$- and ${D}$-meson decay constants from three-flavor lattice
  {QCD}.
\newblock {\em Phys. Rev. D}, 85:114506, Jun 2012.

\bibitem{PhysRevD.91.054507}
N.~Carrasco, P.~Dimopoulos, R.~Frezzotti, P.~Lami, V.~Lubicz, F.~Nazzaro,
  E.~Picca, L.~Riggio, G.~C. Rossi, F.~Sanfilippo, S.~Simula, and C.~Tarantino.
\newblock Leptonic decay constants ${f}_{K}$, ${f}_{D}$, and ${f}_{{D}_{s}}$
  with ${N}_{f}=2+1+1$ twisted-mass lattice {QCD}.
\newblock {\em Phys. Rev. D}, 91:054507, Mar 2015.

\bibitem{2014231}
Wen-Ping Chen, Yu-Chih Chen, Ting-Wai Chiu, Han-Yi Chou, Tian-Shin Guu, and
  Tung-Han Hsieh.
\newblock Decay constants of pseudoscalar ${D}$-mesons in lattice {QCD} with
  domain-wall fermion.
\newblock {\em Physics Letters B}, 736:231 -- 236, 2014.

\bibitem{PhysRevD.86.054510}
Heechang Na, Christine T.~H. Davies, Eduardo Follana, G.~Peter Lepage, and
  Junko Shigemitsu.
\newblock $|{V}_{cd}|$ from ${D}$ meson leptonic decays.
\newblock {\em Phys. Rev. D}, 86:054510, Sep 2012.

\bibitem{PhysRevD.88.014015}
P.~Gelhausen, A.~Khodjamirian, A.~A. Pivovarov, and D.~Rosenthal.
\newblock Decay constants of heavy-light vector mesons from {QCD} sum rules.
\newblock {\em Phys. Rev. D}, 88:014015, Jul 2013.

\bibitem{LUCHA201182}
Wolfgang Lucha, Dmitri Melikhov, and Silvano Simula.
\newblock {OPE}, charm-quark mass, and decay constants of ${D}$ and ${D}_s$
  mesons from {QCD} sum rules.
\newblock {\em Physics Letters B}, 701(1):82 -- 88, 2011.

\bibitem{LUCHA201412}
Wolfgang Lucha, Dmitri Melikhov, and Silvano Simula.
\newblock Decay constants of the charmed vector mesons ${D}^*$ and ${D}_s^*$
  from {QCD} sum rules.
\newblock {\em Physics Letters B}, 735:12 -- 18, 2014.

\bibitem{NARISON20131321}
Stephan Narison.
\newblock A fresh look into $\overline{m}_{c,b} (\overline{m}_{c,b}) $ and
  precise $f_{D_{(s)},B_{(s)}}$ from heavy-light {QCD} spectral sum rules.
\newblock {\em Physics Letters B}, 718(4):1321 -- 1333, 2013.

\bibitem{Wang2015}
Zhi-Gang Wang.
\newblock Analysis of the masses and decay constants of the heavy-light mesons
  with {QCD} sum rules.
\newblock {\em The European Physical Journal C}, 75(9):427, Sep 2015.

\bibitem{Yang2012}
Mao-Zhi Yang.
\newblock Wave functions and decay constants of ${B}$ and ${D}$ mesons in the
  relativistic potential model.
\newblock {\em The European Physical Journal C}, 72(2):1880, Feb 2012.

\bibitem{PhysRevD.99.093002}
Hao-Kai Sun and Mao-Zhi Yang.
\newblock Wave functions and leptonic decays of bottom mesons in the
  relativistic potential model.
\newblock {\em Phys. Rev. D}, 99:093002, May 2019.

\bibitem{EBERT200693}
D.~Ebert, R.N. Faustov, and V.O. Galkin.
\newblock Relativistic treatment of the decay constants of light and heavy
  mesons.
\newblock {\em Physics Letters B}, 635(2):93 -- 99, 2006.

\bibitem{CVETIC200484}
G.~Cveti\v{c}, C.S. Kim, Guo-Li Wang, and Wuk Namgung.
\newblock Decay constants of heavy meson of $0^-$ state in relativistic
  {S}alpeter method.
\newblock {\em Physics Letters B}, 596(1):84 -- 89, 2004.

\bibitem{PhysRevD.81.054022}
Chien-Wen Hwang.
\newblock ${SU}(3)$ symmetry breaking in decay constants and electromagnetic
  properties of pseudoscalar heavy mesons.
\newblock {\em Phys. Rev. D}, 81:054022, Mar 2010.

\bibitem{Becirevic:2014kaa}
Damir Becirevic, Alain Le~Yaouanc, Arantza Oyanguren, Patrick Roudeau, and
  Francesco Sanfilippo.
\newblock {Insight into $D/B\to \pi \ell \nu_\ell$ decay using the pole
  models}.
\newblock 2014.

\bibitem{Lucha:2014hqa}
Wolfgang Lucha, Dmitri Melikhov, and Silvano Simula.
\newblock {{QCD} sum-rule results for heavy-light meson decay constants and
  comparison with lattice {QCD}}.
\newblock In {\em {8th International Workshop on the CKM Unitarity Triangle
  (CKM 2014) Vienna, Austria, September 8-12, 2014}}, 2014.

\bibitem{NARISON2016143}
Stephan Narison.
\newblock Decay constants of heavy-light mesons from {QCD}.
\newblock {\em Nuclear and Particle Physics Proceedings}, 270-272:143 -- 153,
  2016.
\newblock 18th Montpellier International Conference on Quantum Chromodynamics
  (QCD 15).

\bibitem{osti_7335315}
SS~Gershtein and M~Yu Khlopov.
\newblock Leptonic decays of heavy pseudoscalar mesons.
\newblock {\em JETP Lett}, 23(6):338, 1976.

\end{thebibliography}

\end{document}